\documentclass[lettersize,journal]{IEEEtran}

\usepackage{amsmath,amsfonts}
\usepackage{algorithmic}
\usepackage{algorithm}
\usepackage{array}
\usepackage[caption=false,font=normalsize,labelfont=sf,textfont=sf]{subfig}
\usepackage{textcomp}
\usepackage{stfloats}
\usepackage{url}
\usepackage{xurl}
\usepackage{verbatim}
\usepackage{graphicx}
\usepackage{cite}
\usepackage{bbm}
\usepackage{booktabs}
\usepackage{makecell}
\usepackage{siunitx}
\newcolumntype{P}[1]{>{\raggedright\arraybackslash}p{#1}}
\newcolumntype{C}[1]{>{\centering\arraybackslash}p{#1}}

\usepackage[acronym]{glossaries}

\newacronym{uwb}{UWB}{Ultra-Wideband}
\newacronym{ble}{BLE}{Bluetooth Low Energy}
\newacronym{tdoa}{TDoA}{Time Difference of Arrival}
\newacronym{oob}{OOB}{Out-of-Band}
\newacronym{tof}{TOF}{Time-of-Flight}
\newacronym{sts}{STS}{Scrambled Timestamp Sequence}
\newacronym{phy}{PHY}{Physical Layer}
\newacronym{mac}{MAC}{Medium Access Control}
\newacronym{ss}{SS}{Single-Sided}
\newacronym{ds}{DS}{Double-Sided}
\newacronym{twr}{TWR}{Two-Way Ranging}
\newacronym{mic}{MIC}{Message Integrity Check}
\newacronym{pdu}{PDU}{Protocol Data Unit}
\newacronym{ack}{ACK}{Acknowledgement}
\newacronym{ifs}{IFS}{inter-frame space}
\newacronym{prf}{PRF}{Pulse Repetition Frequency}
\newacronym{shr}{SHR}{synchronisation header}
\newacronym{sfd}{SFD}{start frame delimiter}
\newacronym{phr}{PHR}{PHY header}
\newacronym{hrp}{HRP}{High-Rate Pulse}
\newacronym{psr}{PSR}{Preamble Symbol Repetition}
\newacronym{ppdu}{PPDU}{PHY Protocol Data Unit}
\newacronym{psdu}{PSDU}{PHY Service Data Unit}
\newacronym{ul}{UL}{Uplink}
\newacronym{dl}{DL}{Downlink}
\newacronym{cs}{CS}{Channel Sounding}
\newacronym{cir}{CIR}{Channel Impulse Response}
\newacronym{ndp}{NDP}{Neighbor Discovery Process}
\newacronym{ultdoa}{UL-TDoA}{Uplink-Time Difference of Arrival}
\newacronym{dltdoa}{DL-TDoA}{Downlink-Time Difference of Arrival}
\newacronym{dstwr}{DS-TWR}{Double-Sided Two-Way Ranging}
\newacronym{sstwr}{SS-TWR}{Single-Sided Two-Way Ranging}
\newacronym{tx}{TX}{Transmit}
\newacronym{rx}{RX}{Receive}
\newacronym{dp}{DP}{Data Portion}
\newacronym{cte}{CTE}{Constant Tone Extension}
\newacronym{rtls}{RTLS}{Real-Time Location System}
\newacronym{iot}{IoT}{Internet of Things}
\newacronym{ccc}{CCC}{Car Connectivity Consortium}
\newacronym{pi}{PI}{periodic-interval}
\newacronym{csa}{CSA}{Connectivity Standards Alliance}
\newacronym{aoa}{AoA}{Angle of Arrival}

\begin{document}

\title{Analytical Modeling of Latency and Energy Trade-offs in FiRa-Like BLE-UWB Ranging and Localization Systems}

\author{Adelina Giurea,
        Jorg Wieme,
        Ben Van Herbruggen,
        Jeroen Hoebeke,
        Eli De Poorter

\thanks{The authors are with IDLab Research Group, Department of Information Technology, Ghent University--imec, 9052 Ghent, Belgium (e-mail: adelina.giurea@ugent.be).}

}



\maketitle

\begin{abstract}
Accurate position and distance information provided by \gls{uwb} can enhance a wide range of applications. However, its receive-mode power consumption makes continuous channel monitoring impractical for energy-constrained devices. Consequently, \gls{uwb} systems often rely on \gls{ble} to discover peers and establish sessions before ranging begins. Although both technologies have been studied in isolation, an framework capturing the latency and energy consumption of the complete workflow is still lacking. To address this gap, we develop a unified analytical model for \gls{ble}-assisted \gls{uwb} systems, from device-to-device ranging to infrastructure-based localization. The model includes closed-form expressions for discovery, parameter exchange, and ranging, and evaluates the resulting latency and energy consumption. Evaluation of three representative \gls{ble} configurations shows that the scanner consumes 28-57$\times$ more energy per discovery than the advertiser. In addition, the 99th-percentile discovery latency is 2-3.4$\times$ higher than the mean and reaches up to 110 s in the lowest-power configuration. The one-time \gls{ble} setup overhead is amortized after 24-562 ranging rounds, depending on the discovery configuration. Overall, the analysis demonstrates that the \gls{ble} discovery phase can dominate both the latency and the energy budget during short \gls{uwb} interactions, making discovery configuration as important as the ranging scheme.
\end{abstract}


\glsresetall

\section{Introduction}
\label{sec:introduction}
\IEEEPARstart{S}{patial} awareness capabilities are integrated in the majority of mobile consumer devices such as smartphones, smartwatches, and smart door locks. At the same time, automotive manufacturers are equipping their vehicles and keys with wireless ranging to mitigate relay attacks. In large industrial environments, thousands of items are simultaneously tracked wirelessly. All these applications share the same key principles: (secure) high accuracy ranging, optimized radio usage to scale to systems with a large number of tags and low power consumption to enable long lifetime on a single battery. Although these ranging capabilities are integrated into well-established standards, including Bluetooth Channel Sounding \cite{wieme2025cs}, Wi-Fi fine timing measurement \cite{mawifi2022}, and 5G \cite{dwivedipositioning5g2021}, their localization accuracy remains below that achievable with \gls{uwb} impulse radio, which continues to be the preferred technology for high-accuracy positioning. Its large bandwidth (typically 500 MHz or more) enables resolving frame arrival times with picosecond-level precision, supporting centimeter-level ranging even in multipath environments~\cite{performancecomparissondw1000dw3000}. Recent studies have further demonstrated millimeter-level accuracy by exploiting phase information~\cite{ma2025muloc,demoerloose2026mm}. Besides high ranging accuracy, the IEEE 802.15.4z~\cite{ieee802154z} amendment also introduced secure ranging through the addition of the \gls{sts} to the \gls{uwb} packet structure.

While \gls{uwb} excels at accurate ranging and sensing, it is less suited for continuous channel monitoring and low-power device discovery~\cite{uwbnd2026}, as detecting sporadic transmissions requires a receiver to continuously sample a wide bandwidth and process high-rate signals. This incurs a significant energy cost, making always-on \gls{uwb} channel monitoring and contention-based \gls{uwb} medium access impractical. 

To address this limitation, commercial battery-constrained \gls{uwb} systems typically use \gls{ble} as a companion radio to provide low-power device discovery. A \gls{ble} connection can then be used for authentication and the exchange of \gls{uwb} session parameters before ranging starts. Similar principles are applied in smartphone operating systems (Android \cite{androiduwb2026} and iOS \cite{appleuwb2026}) and standardization efforts such as the FiRa specification \cite{firadiscover}. In this work, we refer to systems following this approach as FiRa-like systems.

The hybrid \gls{ble}-\gls{uwb} architecture introduces a trade-off between session setup and ranging. Establishing a session requires a one-time energy and latency cost for \gls{ble} discovery and commissioning, while subsequent ranging rounds add an incremental cost. As a result, the setup overhead can dominate short interactions (e.g. access control or point-and-control) particularly because \gls{ble} discovery is stochastic and strongly parameter-dependent. In infrastructure-based localization, the \gls{uwb} cost can increase further as a single positioning update may involve multiple anchors and ranging exchanges.

Prior work has treated \gls{ble} neighbor discovery and \gls{uwb} ranging as separate components, leaving the end-to-end trade-offs of a complete \gls{ble}-assisted \gls{uwb} session unexplored. In particular, it remains unclear when \gls{ble} setup dominates latency and energy consumption, how many ranging rounds are needed to amortize this overhead, and which phase has the greatest impact on short interactions. Moreover, existing analyses on latency or energy primarily focus on device-to-device ranging, while infrastructure-based localization involving multiple anchors remains largely unexplored. 

\newpage
The main contributions of this work are as follows:
\begin{itemize}
    \item We present, to the best of our knowledge, the first end-to-end analytical model of latency and energy consumption of \gls{ble}-assisted \gls{uwb} sessions, covering \gls{ble} discovery, \gls{ble} parameter negotiation, \gls{uwb} startup, and ranging within a single framework.
    \item We extend the model beyond device-to-device ranging to infrastructure-based localization. The framework covers multi-anchor \gls{twr}, \gls{ultdoa}, and \gls{dltdoa}, capturing the additional communication and ranging phases introduced by each localization scheme.
    \item Using the model, we quantify how \gls{ble} discovery and setup contribute to the overall latency and energy consumption of ranging and localization sessions. We analyze the resulting asymmetry between devices, the growth of latency with the number of anchors or ranging rounds, and the number of ranging rounds required to amortize the \gls{ble} setup overhead.
    \item We provide a public release of our code, enabling researchers and practitioners to estimate the latency and energy consumption for their own parameter configurations.\footnote{Code will be made available upon acceptance.}
\end{itemize}

The remainder of this paper is organized as follows. Section~\ref{sec:related_work} reviews related work. Section \ref{sec:analytical_model} describes the analytical model, while Section \ref{sec:discussions} discusses the model and its results. Section \ref{sec:ranging-to-localization} extends the model to localization, Section \ref{sec:limitations-future-work} outlines the limitations of the model and future work, and Section \ref{sec:conclusion} concludes the paper.

\section{Related Work}
\label{sec:related_work}
As this is the first paper to analyze \gls{ble}-assisted \gls{uwb} systems, the related work is divided into the following parts: i) \gls{ble} neighbor discovery, ii) \gls{uwb} ranging and localization and iii) hybrid multi-radio systems. In the fourth part of this section, current standardization efforts are discussed to position this work.

\subsection{\gls{ble} Neighbor Discovery: Latency and Energy}
\gls{ble} literature primarily studies neighbor discovery, focusing on the latency and energy consumption associated with advertising and scanning. 
Energy-efficient discovery between unsynchronized, duty-cycled devices is a classical problem in mobile computing. In \gls{ble}, discovery is governed by a \gls{pi} scheme defined by the advertising interval, scan interval, and scan window. In~\cite{kindt2018neighbordiscovery} the authors, developed a comprehensive model for \gls{ble}-like protocols, proving that discovery latency is deterministic and bounded for almost all valid parameter configurations. However, their model evaluates discovery only up to the first successful packet reception. 
Existing \gls{ble} research models the discovery process on the three primary advertising channels, the pseudo-random advertising delay, and the packet exchange following a received advertisement. Their main differences are in how they account for the unknown phase offset between the advertiser and scanner and whether the scanner operates continuously or in a duty-cycled way. The study presented in~\cite{liu2012} partitions the three-channel advertising cycle according to where the advertising event starts and averages over the resulting cases. They assume continuous scanning, resulting in higher energy consumption than duty-cycled operation. In addition, the scan window is assumed to be at least as long as the advertising interval, ensuring that every advertising event can be observed within a single scan cycle. While these assumptions simplify the analysis, they do not capture the latency-energy trade-offs introduced by shorter scan windows. The authors in~\cite{cho2015latency} instead derive a per-channel success probability that accounts for the scanner's arrival offset through the residual scan window time, and sum these probabilities over repeated advertising attempts to obtain a closed-form expected latency. The scanner is allowed to duty-cycle, with continuous scanning as the special case. The authors of~\cite{kindt2018neighbordiscovery} extended their model to \gls{ble} in~\cite{kindt2020} where they numerically integrate over the advertiser-scanner offset and approximate the cumulative random advertising delay, supporting both duty-cycled and continuous scanning. They also extend the model to the energy consumption of discovery and subsequent \gls{ble} connection states. The work reported in~\cite{shih2022latency} introduced a discrete timing model that captures 3-channel rotation and identifies latency bounds and synchronizing locks when advertising and scanning intervals align.

Two aspects remain open. First, the interval reserved from the scan window for the packet exchange is defined differently in each work: only the time to receive the \texttt{ADV\_IND} packet~\cite{liu2012}, only the response handshake~\cite{cho2015latency}, or the whole exchange~\cite{shih2022latency}. More importantly, only~\cite{kindt2020} extends the analysis beyond discovery to the \gls{ble} connection state. However, it does not retain the outcome of the discovery process, such as the advertising round and channel on which discovery succeeds or the resulting energy cost for each device role. Existing models also do not distinguish between scan-response and connection-based discovery in the context of a subsequent \gls{uwb} session. In particular, none carry the discovery and connection outcomes into the establishment and ranging phases.
In this work, we build upon the 3-channel timing formulation of~\cite{shih2022latency}, but refine the effective scan window by reserving only the duration until the preamble and access address are received. Furthermore, we extract the exact successful round and channel to map discovery outcomes directly into state-level energy models and subsequent session establishment.

\subsection{\gls{uwb} Ranging and Localization: Latency and Energy}
Most \gls{uwb} literature focuses on ranging and localization accuracy~\cite{ma2025muloc,zafari2019,singleanchorlocalization2022}, with there is little attention paid to latency and energy consumption. In~\cite{flueratoru2020energy}, the authors characterize the energy consumption of different transceiver states and show that reception consumes more energy than transmission since reception requires continuous correlation to search for the signal, while transmission only consumes power during the burst. In addition, they show that higher data rates reduce the overall energy by shortening packet duration, despite higher current consumption. The power consumption of a complete \gls{dstwr} exchange is measured in~\cite{performancecomparissondw1000dw3000}. From these measurements it is concluded that receiver-active intervals between message transmissions dominate energy consumption, with channel, data rate, preamble length, and \gls{prf} as the main drivers of peak power. 

The work presented in~\cite{tiemann2019scalability} models the scalability, real-time performance, and energy efficiency of \gls{twr} and \gls{tdoa}, while the authors of ~\cite{ridolfiscalability2018} study \gls{uwb} localization at high user densities. These works show that the number of messages a device transmits and receives, and hence the time it remains active, is a key factor in localization energy consumption.

However, these studies assume that the devices are already configured for \gls{uwb} ranging, including the required \gls{phy} parameters and, where applicable, schedules and keys. In secure ranging, the \gls{sts} is derived from a session key that must be established before ranging begins~\cite{secureranging2024}. The time and energy required to obtain this configuration are outside the scope of these works. Our model addresses this missing setup phase by linking \gls{ble} discovery and negotiation to the subsequent \gls{uwb} ranging or localization exchange.

\subsection{Hybrid \gls{uwb} and \gls{oob} Systems}
Using a low-power radio to trigger a more power-hungry one is a well-established approach, ranging from wake-up receivers in WLAN~\cite{wakeupreceivers2022} to sub-GHz companion radios in \gls{uwb}~\cite{vanherbruggen2019wipos}. In Wi-PoS~\cite{vanherbruggen2019wipos}, the sub-GHz link handles the communication and reporting, while \gls{uwb} is used for ranging, reducing the reported power consumption by $14.3\%$ at the tag and $20.5\%$ at the anchor node. 
An alternative is to perform discovery directly over \gls{uwb}, avoiding a companion radio altogether. In \gls{uwb}-ND~\cite{uwbnd2026}, advertisers periodically perform a short channel activity detection, while the scanner transmits a long wake-up call. Compared to \gls{pi} discovery, \gls{uwb}-ND reduces tag energy by up to 50\% when discovery requests are rare (long idle windows of $\geq 70$ s), but increases it by up to 64\% when requests are frequent. Moreover, all nodes must share the sniffing interval, timing parameters and wake-up key in advance. Such approaches therefore target networks with mains-powered anchors and preconfigured nodes, rather than discovery between previously unknown devices.

\subsection{Standardization and Ecosystem Trends}
To improve interoperability and secure operation of \gls{uwb} systems, different (industry-driven) standardization bodies are developing their own specifications for certification. These specifications often use a low-power radio for discovery, might use rough distance estimation based on RSSI values, and session setup before the actual \gls{uwb} ranging phase. The IEEE 802.15.4z~\cite{ieee802154z}, introduced in 2020, defines secure ranging, while leaving discovery and session setup for upper layers. The authors in~\cite{coppens2022overview} note that \gls{phy}-level compatibility between \gls{uwb} ecosystems is largely a configuration issue, whereas discovery and upper-layer interoperability remain open. FiRa typically addresses this through an \gls{oob} channel for discovery and session establishment using \gls{ble}~\cite{firausecases}. Besides \gls{twr}, FiRa supports \gls{dltdoa}-based indoor tracking~\cite{firauntracked2023}, and since FiRa Core 4.0, \gls{ultdoa} asset tracking~\cite{firaassettracking}. Other ecosystems follow a similar approach: CCC Digital Key~\cite{ccc2022digitalkey} authenticates phone and vehicle over \gls{ble} before a secure \gls{uwb} ranging session is established, and Aliro 1.0~\cite{csa2026aliro} specifies \gls{ble} in combination with \gls{uwb} for hands-free building access. Android~\cite{androiduwb2026} and Apple~\cite{appleuwb2026} require ranging parameters to be exchanged over an \gls{oob} transport, typically \gls{ble}. Vendor implementations likewise use \gls{ble} for discovery and to trigger \gls{uwb} on demand~\cite{samsungfira2025,nxp2021bleuwb}. The capability request, capability response, and configuration messages defined by Android~\cite{androiduwb2026} provide the basis for the negotiation phase in our model. The diversity of ranging parameters and deployment scenarios makes the negotiation of operational parameters an important component of modern \gls{uwb} systems.

\section{Analytical Model}
\label{sec:analytical_model}
\begin{figure*}
    \centering
    \includegraphics[width=1\linewidth]{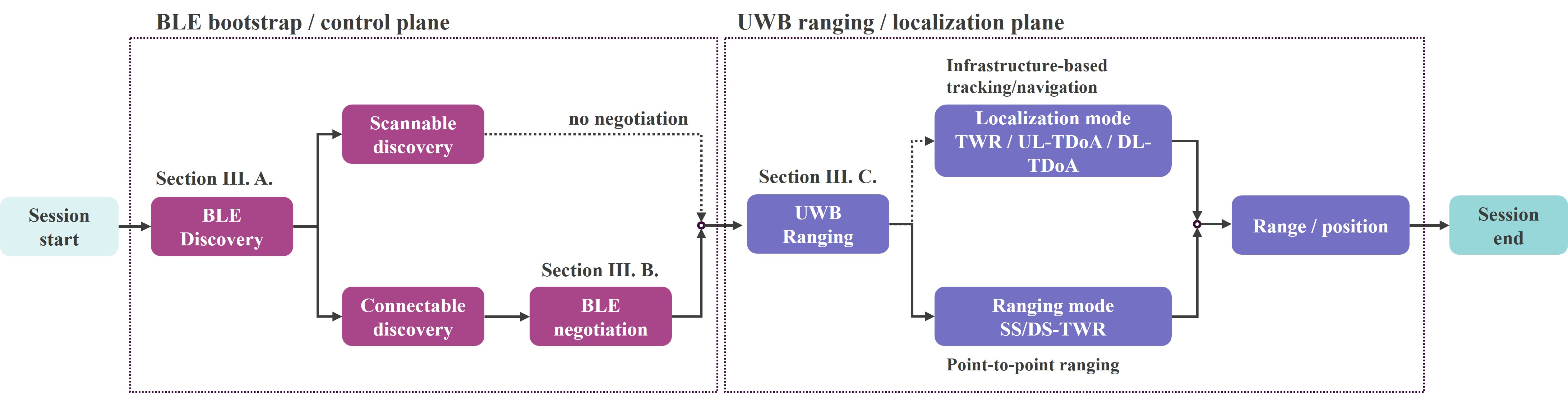}
    \caption{End-to-end model of a \gls{ble}-assisted \gls{uwb} system. \gls{ble} discovery (Section~\ref{subsec:ble_discovery}) followed by parameters exchange (Section~\ref{subsec:neg}). Both converge at \gls{uwb} ranging (Section~\ref{subsec:uwb_ranging}) where it splits into ranging or localization before the session ends.}
    \label{fig:schematic-overview}
\end{figure*}

The goal of the model is to derive the expected end-to-end delay and energy consumption of FiRa-like systems by combining (i) \gls{ble} neighbor discovery latency/energy, (ii) \gls{ble} negotiation overhead, and (iii) \gls{uwb} startup, and ranging overhead. The three phases of the model are illustrated in Fig.~\ref{fig:schematic-overview}.

\subsection{\gls{ble} Neighbor Discovery}
\label{subsec:ble_discovery}
\begin{figure}
    \centering
    \includegraphics[width=1\linewidth]{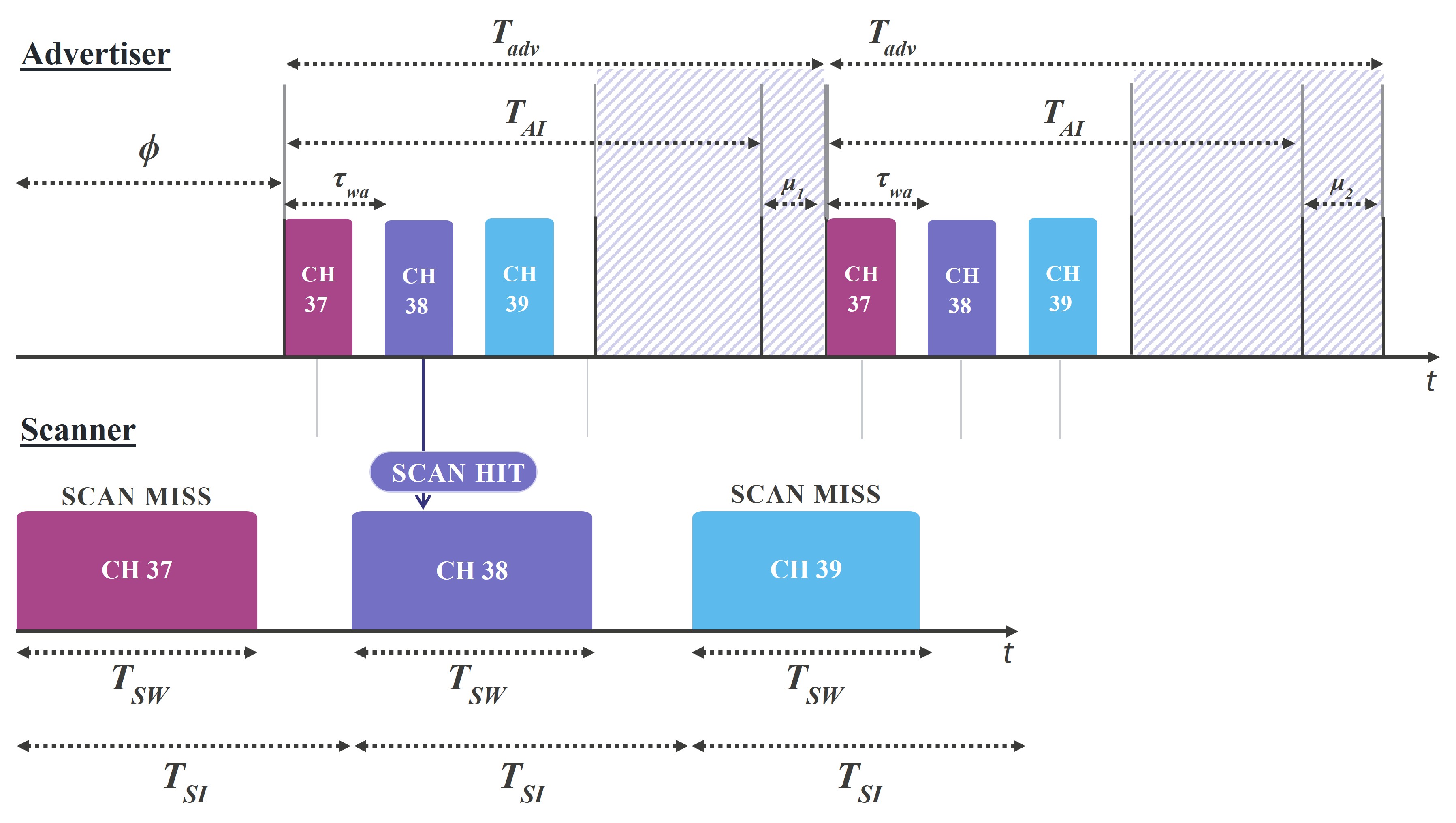}
    \caption{BLE \gls{ndp}: the advertiser
    transmits one \texttt{ADV\_IND} packet per primary channel (37, 38, 39) within each advertising event. Consecutive advertising events are separated by the fixed interval $T_{\mathrm{AI}}$ plus a pseudo-random delay $\mu$ (replaced by $\bar{\mu}$ in the model). The hatched area indicates the interval during which the transmitter sleeps. The scanner listens on one channel per scan interval $T_{\mathrm{SI}}$ for a scan window $T_{\mathrm{SW}}$. Discovery occurs when the scanner's active window overlaps with an advertising event on the same channel.}
    \label{fig:neighbor_discovery}
\end{figure}

\gls{ble} is designed for devices with small batteries that must remain on a low duty cycle. It operates in the 2402--2480 MHz band, divided into 40 channels of 2 MHz. Three of these, channels 37, 38, and 39, are reserved for advertising and carry the \gls{ndp} (illustrated in Fig. \ref{fig:neighbor_discovery}) through which two devices discover each other before any data can be exchanged. The discovery parameters strongly influence discovery latency and power consumption and are listed in Table \ref{tab:ble_discovery_params}, as defined in \cite{bluetooth_core_6.3_ll}.

\begin{table}
    \caption{\gls{ble} Discovery Parameters}
    \label{tab:ble_discovery_params}
\centering
\renewcommand{\arraystretch}{1.3}
\begin{tabular}{lP{4.0cm}l}
\toprule
    \textbf{Symbol} & \textbf{Description} & \textbf{Range/Value}\\
    \midrule
    $T_{\mathrm{AI}}$ & Advertising interval (step size: $0.625\,\mathrm{ms}$) & $20\,\mathrm{ms}$--$10485.76\,\mathrm{s}$ \\
    $\mu$ & Pseudo-random advertising delay & $[0,10]\,\mathrm{ms}$ \\
    $\bar{\mu}, \mu_{\mathrm{max}}$ & Mean/max pseudo-random delay & $5\,\mathrm{ms}$,$10\,\mathrm{ms}$\\
    $T{_{\mathrm{adv}}}$ & Advertising event, $T_{\mathrm{AI}} + {\mu}$ & - \\
    $\bar{T}{_{\mathrm{adv}}}$ & Mean advertising event, $T_{\mathrm{AI}} + \bar{\mu}$ & - \\
    $\tau_{\mathrm{wa}}$ & Time between two consecutive \texttt{ADV\_INDs} & $\le 10\,\mathrm{ms}$ \\
    $T_{\mathrm{SI}}$ & Scan interval & $[2.5\,\mathrm{ms},40.96\,\mathrm{s}]$ \\
    $T_{\mathrm{SW}}$ & Scan window & $[2.5\,\mathrm{ms},T_{\mathrm{SI}}]$\\
    $T_{\mathrm{IFS}}$ & Inter-frame space & $150\,\mathrm{\mu s}$\\
\bottomrule
\end{tabular}
\end{table}

\begin{table}
    \caption{\gls{ble} Packet Format in LE 1M \gls{phy}}
    \label{tab:ble_packet_format}
\centering
\renewcommand{\arraystretch}{1.3}
\begin{tabular}{lP{4.5cm}l}
\toprule
    \textbf{Symbol} & \textbf{Description} & \textbf{Value}\\
    \midrule
    $R_{\mathrm{PHY}}$& \gls{phy} data rate & $1\,\mathrm{Mb/s}$ \\
    $S_{\mathrm{PRE}}$ & Preamble field size & $1\,\mathrm{B}$ \\
    $S_{\mathrm{AA}}$ & Access address field size & $4\,\mathrm{B}$\\
    $S_{\mathrm{HDR}}$ & \gls{pdu} header size & $2\,\mathrm{B}$\\
    $S_{\mathrm{CRC}}$ & CRC field size & $3\,\mathrm{B}$\\
    $S_{\mathrm{MIC}}$ & Message Integrity Check field size & $0/4\,\mathrm{B}$ \\
    $S_{\mathrm{ADV}}$ & \texttt{ADV\_IND} payload size & $6$--$37\,\mathrm{B}$ \\
    $S_{\mathrm{REQ}}$ & \texttt{SCAN\_REQ} payload size & $12\,\mathrm{B}$ \\
    $S_{\mathrm{RSP}}$ & \texttt{SCAN\_RSP} payload size & $6$--$37\,\mathrm{B}$ \\
    $S_{\mathrm{CONN}}$ & \texttt{CONNECT\_IND} payload size & $34\,\mathrm{B}$ \\
\bottomrule
\end{tabular}
\end{table}

The \textit{advertiser} periodically advertises \texttt{ADV\_IND} packets to announce its presence and capabilities. A transmission across all three channels forms an advertising event $T_{\mathrm{adv}}$, with $\tau_{\mathrm{wa}}$ denoting the time between consecutive \texttt{ADV\_IND} packets. Advertising events repeat every $T_{\mathrm{adv}}=T_{\mathrm{AI}}+\mu$, where $T_{\mathrm{AI}}$ is the advertising interval and $\mu$ is a pseudo-random delay that reduces repeated packet collisions by preventing advertisers from remaining in sync.

The \textit{scanner} operates independently. It turns its receiver on for a fixed scan window $T_{\mathrm{SW}}$ once every scan interval $T_{\mathrm{SI}}$, then moves to the next advertising channel and listens again. Because the advertiser and scanner run on separate, unsynchronized clocks, discovery succeeds only when a scan window happens to overlap an \texttt{ADV\_IND} packet. This timing overlap, adapted from~\cite{shih2022latency}, is formalized in the model below and determines the discovery latency.

\subsubsection{Timing}

To derive the latency and energy consumption of the \gls{ble} procedures, the on-air time of each exchanged packet is computed as 
\begin{equation}
    S_{\mathrm{PDU}} = S_{\mathrm{HDR}} + S_{\mathrm{TYPE}} +  S_{\mathrm{MIC}},
\end{equation}
\begin{equation} 
T_{\mathrm{{TYPE}}} = \frac{8\times(S_{\mathrm{PRE}} + S_{\mathrm{AA}} + S_{\mathrm{CRC}} + S_{\mathrm{PDU}})}{R_{\mathrm{PHY}}}, 
\label{eq:tpkt} \end{equation} 
where $S_{\mathrm{MIC}} = 4\,\mathrm{B}$ on encrypted data packets and 0 otherwise. $S_{\mathrm{TYPE}}$ is the payload size of packet type $\mathrm{TYPE}\in\{\mathrm{ADV},\mathrm{REQ},\mathrm{RSP},\mathrm{CONN}\}$ in bytes and $R_{\mathrm{PHY}}$ is the \gls{phy} data rate. All parameter values are listed in Table~\ref{tab:ble_packet_format}.

\subsubsection{Discovery Latency}

With the on-air time defined, we can now describe the discovery timing itself. 

Let $t=0$ be the start of a scan interval on channel 37, and let $\phi$ denote the offset between that instant and the first \texttt{ADV\_IND} packet transmitted on channel 37. Since the advertiser and scanner are unsynchronized, $\phi$ lies in the interval $[0,3T_{\mathrm{SI}})$. To simplify the model, \cite{shih2022latency} replaced the pseudo-random delay $\mu$ by its mean 
\begin{equation}
    \bar{T}_{\mathrm{adv}} = T_{\mathrm{AI}} + \bar{\mu},
    \qquad
    \bar{\mu} = \tfrac{1}{2}\mu_{\mathrm{\max}} = 5~\mathrm{ms},
    \label{eq:mean_adv_interval}
\end{equation}
where $\mu \sim \mathcal{U}[0,\mu_{\mathrm{\max}}]$ and $\mu_{\mathrm{\max}}=10$~ms. Indexing the advertising channels as $c_1=37$, $c_2=38$, $c_3=39$, the transmission start of the \texttt{ADV\_IND}
packet sent on channel $c_k$ during advertising round $r$ is then

\begin{equation}
\begin{split}
     t_{k,r}
    & =
    \phi + (k-1)\,\tau_{\mathrm{wa}} + r\,\bar{T}_{\mathrm{adv}},
    \\ \qquad k & \in \{1,2,3\},\; r \in \mathbb{N}_{0}.
    \label{eq:adv_epoch}
\end{split}
\end{equation}

The scanner visits channels $c_1, c_2, c_3$ cyclically with period $3T_{\mathrm{SI}}$, scanning channel $c_k$ during $[(k-1)T_{\mathrm{SI}},\,(k-1)T_{\mathrm{SI}}+T_{\mathrm{SW}})$ of every cycle. Reception succeeds only if the receiver synchronizes to the packet, i.e. receives its preamble and access address, before it is switched off. Once synchronized, the receiver is assumed to complete the reception even if it extends beyond the scan window. The transmission must therefore start at least
$T_{\mathrm{sync}} = 8\times(S_{\mathrm{PRE}} + S_{\mathrm{AA}})/R_{\mathrm{PHY}}$
before the end of the window as in \eqref{eq:eff_window}.

\begin{equation}
    T_{\mathrm{SW}}^{\mathrm{eff}} = T_{\mathrm{SW}} - T_{\mathrm{sync}},
    \label{eq:eff_window}
\end{equation}

which is non-empty provided $T_{\mathrm{sync}} < T_{\mathrm{SW}} \le T_{\mathrm{SI}}$. Packet $(k,r)$ is therefore received if 
\begin{equation}
    \begin{split}
        t_{k,r} \bmod 3T_{\mathrm{SI}}
          & \in
        \bigl[(k-1)T_{\mathrm{SI}},\;
              (k-1)T_{\mathrm{SI}} + T_{\mathrm{SW}}^{\mathrm{eff}}\bigr],
        \qquad \\ k & \in \{1,2,3\},
    \end{split}
    \label{eq:success_cond}
\end{equation}

with $t_{k,r}$ as in \eqref{eq:adv_epoch}. The instants $t_{k,r}$ are strictly increasing under the lexicographic order $(r,k)$, so the packets are enumerated unambiguously in time. Let $\mathcal{S}(\phi)$ denote the set of index pairs satisfying \eqref{eq:success_cond} for a given offset $\phi$. The first successful reception is then
\begin{equation}
    \bigl(R(\phi),K(\phi)\bigr)
    =
    \operatorname*{lex\,min}_{(r,k)\,\in\,\mathcal{S}(\phi)} (r,k),
    \label{eq:first_success}
\end{equation}
where $R(\phi)$ is the advertising round in which overlap first occurs ($R(\phi)=0$ denoting the first event with offset $\phi$) and $K(\phi)$ the channel index of that packet ($K(\phi)=1$ denoting channel~37).

The advertiser discovery latency, measured from the first transmission of the \texttt{ADV\_IND} packet until reception by the scanner is
\begin{equation}
    L_{\mathrm{adv}}(\phi)
    = (K(\phi) - 1)\tau_{\mathrm{wa}} + R(\phi)\,\bar{T}_{\mathrm{adv}} + T_{\mathrm{ADV}}.
    \label{eq:ble_latency_adv}
\end{equation}

Measured from the start of scanning, the scanner discovery latency is
\begin{equation}
    L_{\mathrm{scan}}(\phi)
    = \phi + L_{\mathrm{adv}}(\phi).
    \label{eq:ble_latency_scan}
\end{equation}

Since $\phi$ is unknown, the expected discovery latencies are obtained by integrating over all its possible values,
\begin{equation}
    \mathbb{E}\!\left[L_x\right]
    = \frac{1}{3T_{\mathrm{SI}}}
      \int_{0}^{3T_{\mathrm{SI}}} L_x(\phi)\,\mathrm{d}\phi,
    \qquad x \in \{\mathrm{adv},\mathrm{scan}\},
    \label{eq:expected_ble_latency}
\end{equation}
where we assume that $\bar{T}_{\mathrm{adv}} \neq nT_{\mathrm{SI}}$.

We extend the model in~\cite{shih2022latency} by making a distinction between two discovery modes. Once a scanner has received an \texttt{ADV\_IND} packet, the response depends on the required \gls{uwb} session type, which determines which of two discovery modes applies and, correspondingly, whether the scanner remains a \textit{scanner} or becomes an \textit{initiator}:
\begin{itemize}
    \item \textbf{Scannable discovery}: the \textit{scanner} sends a \texttt{SCAN\_REQ} and the advertiser replies with a \texttt{SCAN\_RSP} that carries basic \gls{uwb} configuration (e.g. channel, anchor group identifier). The advertiser continues advertising on all three channels regardless of whether a \texttt{SCAN\_REQ} was received, and no \gls{ble} connection is formed. This mode is sufficient when the device only needs to identify which \gls{uwb} infrastructure to join, e.g. when entering a public \gls{tdoa} localization zone where no negotiation or further device authentication is required.
    \item \textbf{Connectable discovery:} the scanner acts as an initiator and sends a \texttt{CONNECT\_IND} in response to an \texttt{ADV\_IND} on the overlapping channel. The advertiser then closes the advertising event and both devices transition to the \textit{Connection} state. The resulting \gls{ble} connection carries the parameter exchange described in Section~\ref{subsec:neg}. This mode is required when security material (session keys, two-way security certificates, configurations, addresses, etc.) must be negotiated before \gls{uwb} ranging begins.
\end{itemize}

Both modes complete the discovery with a short deterministic exchange that follows the received \texttt{ADV\_IND}, separated by the mandatory inter-frame space $T_{\mathrm{IFS}}=150~\mu\mathrm{s}$ \cite{bluetooth_core_6.3_ll}. In scannable discovery the scanner sends a \texttt{SCAN\_REQ} and the advertiser answers with a \texttt{SCAN\_RSP}, in connectable discovery the initiator sends a single \texttt{CONNECT\_IND}:
\begin{equation}
\begin{split}
    \Delta_{\mathrm{sc}} & = 2T_{\mathrm{IFS}} + T_{\mathrm{REQ}}
                           + T_{\mathrm{RSP}}, \\
    \qquad
    \Delta_{\mathrm{conn}} &= T_{\mathrm{IFS}} + T_{\mathrm{CONN}}.
    \label{eq:mode_overhead}
\end{split}
\end{equation}
Since $\Delta_m$, $m\in\{\mathrm{sc},\mathrm{conn}\}$, is independent of $\phi$, we calculate the on-air time using \eqref{eq:tpkt}, this then adds a constant to \eqref{eq:ble_latency_adv}--\eqref{eq:ble_latency_scan} and factors out of \eqref{eq:expected_ble_latency},
\begin{equation}
    \mathbb{E}\!\left[L^{m}_{x}\right]
    = \mathbb{E}\!\left[L_{x}\right] + \Delta_{m},
    \qquad x \in \{\mathrm{adv},\mathrm{scan}\}.
    \label{eq:expected_ble_latency_mode}
\end{equation}
\subsubsection{Energy Consumption}
We adopt the state-level decomposition of an advertising event from \cite{luo2019energy} and \cite{liu2013energy}, and extend it to cover the scannable and connectable discovery modes, distinguishing advertiser- from scanner-side energy consumption. The parameters and corresponding energy consumption used in the discovery phase are summarized in Table~\ref{tab:ble_energy_params}.
To calculate the energy consumption for the advertiser, we distinguish between an unsuccessful and a successful event. For each advertising channel, the advertiser transmits an \texttt{ADV\_IND}, waits for $T_{\mathrm{IFS}}$ switches to RX and opens a receive window, the duration of this is:
\begin{equation}
    T_{\mathrm{rx}}^{m}
    =
    \begin{cases}
    T_{\mathrm{IFS}} + T_{\mathrm{REQ}}, & m=\mathrm{sc},\\
    T_{\mathrm{IFS}} + T_{\mathrm{CONN}}, & m=\mathrm{conn}.
    \end{cases}
    \label{eq:trx}
\end{equation}
which is opened whether or not a packet arrives.
The energy per channel is 
\begin{align}
  E^{\mathrm{ch}}_{\mathrm{adv}} &= P_{\mathrm{tx}}T_{\mathrm{ADV}}
                     + E_{\mathrm{tx\text{-}rx}}
                     + P_{\mathrm{rx}}T_{\mathrm{rx}},
  \label{eq:eslot}
\end{align}
and the per advertising event overhead is 
\begin{equation}
\begin{split}
     E_{\mathrm{evt}} = &E_{\mathrm{wake}}+  E_{\mathrm{xtal}} + E_{\mathrm{stnd}} +  \\
    &E_{\mathrm{start}} + E_{\mathrm{post}}.\\
\end{split}
\end{equation}
The idle period after all advertisements are sent is calculated as in \eqref{eq:idle_period}.
 \begin{equation}
     E_{\mathrm{idle,adv}} = P_{\mathrm{idle}}\bigl(\bar{T}_{\mathrm{adv}}
  - 3(T_{\mathrm{ADV}} + T_{\mathrm{rx}})).
  \label{eq:idle_period}
 \end{equation}

An event in which no \texttt{SCAN\_REQ} or \texttt{CONNECT\_IND} is received costs
\begin{equation}
  E_{\mathrm{adv,no}} = E_{\mathrm{evt}} + 3E^{\mathrm{ch}}_{\mathrm{adv}}
  + 2E_{\mathrm{ch}} + E_{\mathrm{idle,adv}}.
  \label{eq:eadvno}
\end{equation}

Let $K(\phi)$ be the channel index on which discovery occurs. The two modes differ in how the event terminates: in scannable discovery the advertiser replies with a \texttt{SCAN\_RSP} and completes the event on all three channels, whereas in connectable discovery it closes the event immediately after the \texttt{CONNECT\_IND} and enters the \textit{Connection} state. With
\begin{equation}
  N^{(m)}\bigl(K\bigr) =
  \begin{cases}
    3, & m = \mathrm{sc},\\
    K, & m = \mathrm{conn},
  \end{cases}
  \label{eq:nm}
\end{equation}
and exchange cost
\begin{equation}
  E^{(m)}_{\mathrm{ex,adv}} =
  \begin{cases}
    E_{\mathrm{rx\text{-}tx}} + P_{\mathrm{tx}}T_{\mathrm{RSP}},
      & m = \mathrm{sc},\\
    0, & m = \mathrm{conn},
  \end{cases}
  \label{eq:exadv}
\end{equation}

the successful event costs
\begin{equation}
\begin{split}
     E^{(m)}_{\mathrm{adv,succ}}(K) &= E_{\mathrm{evt}}
  + N^{(m)}(K)\,E^{\mathrm{ch}}_{\mathrm{adv}}\\
  &+ \bigl(N^{(m)}(K)-1\bigr)E_{\mathrm{ch}}
  + E^{(m)}_{\mathrm{ex,adv}}
\end{split}
  \label{eq:eadvsucc}
\end{equation}

Since $R(\phi)$ denotes the index of the successful round ($R=0$ for the first event), exactly $R(\phi)$ events fail beforehand:
\begin{equation}
  E^{(m)}_{\mathrm{adv}}(\phi) = R(\phi)\,E_{\mathrm{adv,no}}
  + E^{(m)}_{\mathrm{adv,succ}}\bigl(K(\phi)\bigr).
  \label{eq:eadvphi}
\end{equation}

The scanner is decomposed in the same way. Its per-interval overhead is the same $E_{\mathrm{evt}}$, and an interval in which no \texttt{ADV\_IND} is received consists of listening for the full scan window and staying idle in a low-power mode for the remainder of the scan interval,
\begin{equation}
  E_{\mathrm{idle,scan}} = P_{\mathrm{idle}}
  \bigl(T_{\mathrm{SI}} - T_{\mathrm{SW}}\bigr),
  \label{eq:eidlescan}
\end{equation}
so that
\begin{equation}
  E_{\mathrm{scan,no}} = E_{\mathrm{evt}}
  + P_{\mathrm{rx}}T_{\mathrm{SW}}
  + E_{\mathrm{idle,scan}} .
  \label{eq:escanno}
\end{equation}

In the interval in which discovery occurs the scanner listens only until the \texttt{ADV\_IND} has been received, after which it performs the mode-dependent exchange and leaves the discovery phase. The elapsed listening time in that window is
\begin{equation}
\begin{split}
    \delta(\phi) &= \bigl(L_{\mathrm{scan}}(\phi) \bmod 3T_{\mathrm{SI}}\bigr)
  - \bigl(K(\phi)-1\bigr)T_{\mathrm{SI}},\\
  &\qquad T_{\mathrm{ADV}} < \delta(\phi) \le T_{\mathrm{SW}},  
\end{split}
  \label{eq:delta}
\end{equation}

and the exchange costs
\begin{equation}
  E^{(m)}_{\mathrm{ex,scan}} =
  \begin{cases}
    \begin{aligned}[t]
      &E_{\mathrm{rx\text{-}tx}} + P_{\mathrm{tx}}T_{\mathrm{REQ}}\\
      &\quad{}+E_{\mathrm{tx\text{-}rx}}
        + P_{\mathrm{rx}}T_{\mathrm{RSP}},
    \end{aligned} & m = \mathrm{sc},\\
   E_{\mathrm{rx\text{-}tx}} + P_{\mathrm{tx}}T_{\mathrm{CONN}}, & m = \mathrm{conn}.
  \end{cases}
  \label{eq:exscan}
\end{equation}

The successful scan interval is therefore
\begin{equation}
  E^{(m)}_{\mathrm{scan,succ}}(\phi) = E_{\mathrm{evt}}
  + P_{\mathrm{rx}}\,\delta(\phi)
  + E^{(m)}_{\mathrm{ex,scan}} .
  \label{eq:escansucc}
\end{equation}
Before that interval the scanner has completed $N_{\mathrm{scan}}(\phi) = \bigl\lfloor L_{\mathrm{scan}}(\phi)/T_{\mathrm{SI}} \bigr\rfloor$ unsuccessful ones, so that, analogously to \eqref{eq:eadvphi},
\begin{equation}
  E^{(m)}_{\mathrm{scan}}(\phi) = N_{\mathrm{scan}}(\phi)\,E_{\mathrm{scan,no}}
  + E^{(m)}_{\mathrm{scan,succ}}(\phi).
  \label{eq:escanphi}
\end{equation}

As with the latency, the discovery energy depends on the unknown offset $\phi$. The expected discovery energy is therefore obtained by averaging over $\phi$,
\begin{equation}
\mathbb{E}\!\left[E_x^{(m)}\right] =
\frac{1}{3T_{\mathrm{SI}}} \int_0^{3T_{\mathrm{SI}}} E_x^{(m)}(\phi)\,\mathrm{d}\phi,
\qquad x \in \{\mathrm{adv},\mathrm{scan}\},
\label{eq:expected_energy_discovery}
\end{equation}
in the same way as the latency in~\eqref{eq:expected_ble_latency}.

\begin{table}
    \caption{\gls{ble} power and energy parameters for the nRF54L15 at 1~Mbps, 0~dBm, and 3~V~\cite{nrf54datasheet}.}
    \label{tab:ble_energy_params}
\centering
\renewcommand{\arraystretch}{1.3}
\begin{tabular}{lP{3.0cm}lll}
\toprule
    \textbf{Symbol} & \textbf{Description} & \textbf{Power} & \textbf{Time} & \textbf{Energy} \\
    \midrule
    $P_{\mathrm{rx}}$ & Receive power consumption & 10.2 mW & - & -  \\
    $P_{\mathrm{tx}}$ & Transmit power consumption & 14.4 mW & - & -\\
    $P_{\mathrm{idle}}$ & Idle power consumption & 8.7 $\mu$W & - & - \\
    \midrule   
    $E_{\mathrm{wake}}$ & Waking up the device from sleep & 12.9 mW  & 123 $\mu$s & 1.59 $\mu \mathrm{J}$  \\
    $E_{\mathrm{xtal}}$ & Crystal ramp up & 1.5 mW & 500 $\mu$s & 0.75 $\mu \mathrm{J}$ \\
    $E_{\mathrm{stnd}}$ & Time to standby & 0.6 mW & 1148 $\mu$s & 0.69 $\mu \mathrm{J}$ \\
    
    $E_{\mathrm{start}}$ & Starting the radio & 7.2 mW & 103 $\mu$s &  0.74 $\mu \mathrm{J}$ \\
    $E_{\mathrm{rx\text{-}tx}}$ & Switching from RX to TX & 3.6 mW & 143 $\mu$s & 0.51 $\mu \mathrm{J}$  \\
    $E_{\mathrm{tx\text{-}rx}}$ & Switching from TX to RX & 3.6 mW & 143 $\mu$s & 0.51 $\mu \mathrm{J}$  \\
    $E_{\mathrm{post}}$ & Post-processing time & 8.4 mW & 115 $\mu$s & 0.97 $\mu \mathrm{J}$  \\
    $E_{\mathrm{ch}}$ & Channel switching & 9.9 mW & 62 $\mu$s & 0.61 $\mu \mathrm{J}$ \\ 
\bottomrule
\end{tabular}
\end{table}

\subsection{\gls{ble} Negotiation and Security Exchange}
\label{subsec:neg}
The previous phase ends when the advertiser receives the \texttt{CONNECT\_IND} packet. Both devices then enter the \textit{Connection} state, the \textit{advertiser} becoming the \textit{peripheral} and the \textit{initiator} the \textit{central}. The connection is established when a packet is received \cite{bluetooth_core_6.3_ll}. Because the connection was created with \texttt{CONNECT\_IND} on the primary advertising physical channel, the LE 1M \gls{phy} is used in both directions. We do not consider the \gls{phy} update procedure.

\texttt{CONNECT\_IND} carries transmitWindowDelay $D$, transmitWindowOffset $O$ and transmitWindowSize $W$, which bound the interval in which the \textit{central} sends its first data packet. This window allows the \textit{central} to schedule connection events across concurrent connections. The first transmission from the \textit{central} defines the anchor point, after which connection events recur every connInterval $C$.
\subsubsection{Per-Message Payload Sizes}
\label{subsubsec:negotiation-sizes}
Typical \gls{uwb} initialization requires the device address, session ID, \gls{sts} session key, ranging role and configuration profile. A complete configuration would require many \gls{phy}- and \acrshort{mac}-layer parameters (preamble length, data rate, timing settings, and ranging method (e.g. \gls{ss}- or \gls{ds}-\gls{twr})). Predefined profiles reduce this to a profile identifier plus a small set of session-specific fields. Android~\cite{androiduwb2026} follows a similar approach through the use of a \gls{uwb} configuration type together with a \texttt{sessionKeyInfo} of 64 bits for static \gls{sts} and 128 or 256 bits for provisioned \gls{sts}. Our model therefore assumes only the profile identifier and session-specific parameters are exchanged.

\subsubsection{Message Flow}
\label{subsubsec:message-flow}
Following Android\cite{androiduwb2026}, three messages precede ranging: the \textit{central} first requests the parameter values the \textit{peripheral} supports, the \textit{peripheral} replies with its supported set, and the \textit{central} concludes by selecting one value per parameter. The parameter values used throughout this subsection and their corresponding descriptions are listed in Table~\ref{tab:negotiation_params}.

Let $S_i$ denote the payload of message $i\in\{1,2,3\}$ and let
$S_{\mathrm{opcode}}=1$~B be the message-type field present in all three.

\paragraph{$M_1$: Ranging Capability Request (C\,$\rightarrow$\,P)}
Carries only the message type and a bitmap of the ranging technologies for which capabilities are requested
\begin{equation}
  S_1 = S_{\mathrm{opcode}} + S_{\mathrm{tech}} = 2~\mathrm{B}.
  \label{eq:s1}
\end{equation}

\paragraph{$M_2$: Ranging Capability Response (P\,$\rightarrow$\,C)}
Reports the supported \gls{uwb} parameter sets.
\begin{equation}
\begin{split}
     S_2 &= S_{\mathrm{opcode}} + S_{\mathrm{ch}} + S_{\mathrm{pre}}
      + S_{\mathrm{slot}} + S_{\mathrm{int}}
      + S_{\mathrm{prof}}^{\ast} + S_{\mathrm{role}}^{\ast}\\
      &= 11\,\mathrm{B}.
\end{split}
  \label{eq:s2}
\end{equation}

\paragraph{$M_3$: Ranging Configuration (C\,$\rightarrow$\,P)}
Carries the selected value of each parameter together with the session ID ($S_{\mathrm{sid}}$) and \gls{sts} ($S_{\mathrm{key}}$):
\begin{equation}
  S_3
      = S_{\mathrm{opcode}} + S_{\mathrm{aa}} + S_{\mathrm{sid}} + S_{\mathrm{key}}
      + S_{\mathrm{role}} + S_{\mathrm{prof}}.
  \label{eq:s3}
\end{equation}
This is the dominant message, and the only one whose size depends on the selected \gls{sts} mode, giving $S_3 \in [19,43]$~B. In the worst case $\max_i S_i = S_3 = 43~\mathrm{B} \ll S_{\mathrm{payload,max}} = 250$~B, so
each message fits in a single packet and the exchange consists of exactly three \gls{ble} packets. Ranging starts directly after the last packet, so no reply for the peripheral needed.
\subsubsection{Packet Airtime}
\label{subsubsec:neg-airtime}

The airtime of message $i$ is given by \eqref{eq:tpkt} with $S_\mathrm{TYPE} = S_i$ and $S_{MIC} = 4$~B, since the link is assumed encrypted. The total exchange therefore occupies $\sum_i T_{\mathrm{S_i}} + 2T_{\mathrm{IFS}} \in [0.89,1.08]$~ms over the
considered payload range, which is well below the minimum connection interval.

\subsubsection{Negotiation Latency}
\label{subsubsec:neg-latency}

Measured from the end of the \texttt{CONNECT\_IND} packet to the end of $M_3$, i.e.\ the instant at which the \textit{peripheral} starts ranging, the latency is identical for both roles:
\begin{equation}
  L_{\mathrm{neg}} = T_{\mathrm{conn}} + \sum_{i=1}^{3}T_{\mathrm{S_i}}
                   + 2T_{\mathrm{IFS}} + T_{\mathrm{post}},
  \label{eq:neg-latency}
\end{equation}
where the connection establishment delay follows from
\begin{equation}
  T_{\mathrm{conn}} = D + O + \tfrac{W}{2} + T_{\mathrm{wake}},
  \label{eq:tconn}
\end{equation}
assuming the \textit{central} transmits uniformly within the window.

\subsubsection{Energy Consumption}
\label{subsubsec:neg-energy}
The parameters used throughout this section are summarized in Table~\ref{tab:negotiation_params}. The \textit{central} transmits $M_1$ and $M_3$ and receives $M_2$, while the \textit{peripheral} does the opposite. In addition, the \textit{central} stays idle in a low-power mode for the whole interval $D+O+W/2$ and then transmits, whereas the \textit{peripheral} must open its receive window at $D+O$ and listen for $W/2$ on average before the anchor point is detected.
Hence
\begin{equation}
  \begin{split}
    E_{\mathrm{neg}}^{cen} =\;
      &P_{\mathrm{idle}}\bigl(D+O+\tfrac{W}{2}\bigr) + E_{\mathrm{wake}} + \\ 
      & P_{\mathrm{tx}}\bigl(T_{\mathrm{S_1}} + T_{\mathrm{S_3}}\bigr) + P_{\mathrm{rx}}T_{\mathrm{S_2}} + \\
      &2P_{\mathrm{idle}}T_{\mathrm{IFS}} + E_{\mathrm{post}},
  \end{split}
  \label{eq:neg-energy-central}
\end{equation}
\begin{equation}
  \begin{split}
    E_{\mathrm{neg}}^{per} =\;
      &P_{\mathrm{idle}}\bigl(D+O\bigr) + E_{\mathrm{wake}}
       + P_{\mathrm{rx}}\tfrac{W}{2}\\
      &+ P_{\mathrm{rx}}\bigl(T_{\mathrm{S_1}}+T_{\mathrm{S_3}}\bigr)
       + P_{\mathrm{tx}}T_{\mathrm{S_2}}\\
      &+ 2P_{\mathrm{idle}}T_{\mathrm{IFS}} + E_{\mathrm{post}}.
  \end{split}
  \label{eq:neg-energy-peripheral}
\end{equation}
The message asymmetry compounds the role asymmetry: the \textit{peripheral} not only pays the window-listening penalty $P_{\mathrm{rx}}W/2$, it also receives
the largest packet while transmitting only the middle one. The commissioning energy is therefore structurally unequal between the two roles even though the latency is not.

\begin{table}
    \caption{\gls{ble} Negotiation / Commissioning Parameters
    \label{tab:negotiation_params}
        }
\centering
\renewcommand{\arraystretch}{1.3}
\begin{tabular}{llP{2cm}}

\toprule \textbf{Symbol} & \textbf{Description} & \textbf{Value/Size}\\
\midrule
    $D$ & Transmit window delay & $1.25\,\mathrm{ms}$ \\
    $O$ & Transmit window offset & $[0,C]\,\mathrm{ms}$ \\
    $W$ & Transmit window size & $[1.25,\, \min(10,\,\allowbreak C - 1.25)]\,\mathrm{ms}$ \\
    $C$ & Connection interval & $[7.5\,\mathrm{ms},4\,\mathrm{s}]$ \\
    \midrule
    $S_{\mathrm{opcode}}$ & Message type field & $1\,\mathrm{B}$ \\
    $S_{\mathrm{tech}}$ & Supported technologies & $1\,\mathrm{B}$ \\
    $S_{\mathrm{ch}}$ & Supported channel bitmap & $1\,\mathrm{B}$ \\
    $S_{\mathrm{pre}}$ & Supported preamble indices 9-32 & $4\,\mathrm{B}$ \\
    $S_{\mathrm{slot}}$ & Minimum slot duration & $1\,\mathrm{B}$ \\
    $S_{\mathrm{int}}$ & Minimum supported ranging interval & $2\,\mathrm{B}$ \\
    $S_{\mathrm{prof}}^{\ast}$ & Supported profiles & $1\,\mathrm{B}$ \\
    $S_{\mathrm{role}}^{\ast}$ & Supported role sets & $1\,\mathrm{B}$ \\
    $S_{\mathrm{aa}}$ & \gls{uwb} device address & $4\,\mathrm{B}$\\
    $S_{\mathrm{sid}}$ & Session ID size & $4\,\mathrm{B}$\\
    $S_{\mathrm{key}}$ & \gls{sts} key size & $\{8,16,32\}\,\mathrm{B}$\\
    $S_{\mathrm{role}}$ & Chosen \gls{uwb} role & $1\,\mathrm{B}$\\
    $S_{\mathrm{prof}}$ & Chosen \gls{uwb} profile identifier & $1\,\mathrm{B}$\\                 
\bottomrule
\end{tabular}
\end{table}

\subsection{\gls{uwb} Ranging}
\label{subsec:uwb_ranging}
Following the \gls{ble} negotiation phase, both devices activate and configure their \gls{uwb} subsystems using the exchanged session parameters. 
The parameters used throughout this section are summarized in Table~\ref{tab:uwb_params}.

\begin{table}
    \caption{\gls{uwb} energy parameters for the DW3000 on CH5, 3~V~\cite{dw3000usermanual}.
    \label{tab:uwb_params}
        }
\centering \renewcommand{\arraystretch}{1.3}
\begin{tabular}{lP{4.5cm}l}
\toprule \textbf{Symbol} & \textbf{Description} & \textbf{Value}\\
    \midrule
    $P_{\mathrm{OSC}}$ & Oscillator start-up power consumption & $4.8\,\ \mathrm{mW}$\\
    $P_{\mathrm{IDLE\_PLL}}$ & Idle state with PLL enabled (124.8 MHz clock) power consumption & $54\,\ \mathrm{mW}$\\
    $P^{\mathrm{UWB}}_{\mathrm{tx}}$ & Transmit power consumption & $42\,\ \mathrm{mW}$ \\
    $P^{\mathrm{UWB}}_{\mathrm{rx}}$ & Receive power consumption & $48\,\ \mathrm{mW}$ \\
    $P_{\mathrm{deep\_sleep}}$ & Deep sleep power consumption & $0.78\,\ \mathrm{\mu W}$ \\
    \midrule
    $T_{\mathrm{OSC}}$ & Oscillator start-up interval & $1000\,\ \mathrm{\mu s}$\\
    $T_{\mathrm{IDLE\_PLL}}$ & Idle state with PLL enabled  & $20\,\ \mathrm{\mu s}$\\
    $T_{\mathrm{deep\_sleep}}$ & Idle time between ranging rounds  & variable\\
                       
\bottomrule
\end{tabular}
\end{table}

\subsubsection{Startup}
Before any frame can be transmitted or received, the transceiver must first wake up and prepare its clock and RF synthesizer. This wake-up phase consists of the oscillator start-up interval $T_{\mathrm{OSC}}$ and an idle interval $T_{\mathrm{IDLE\_PLL}}$, during which the RF synthesizer locks and the transceiver becomes ready to transmit or receive. The time needed is

\begin{equation} 
    T_{\mathrm{start}}^{\mathrm{UWB}} = T_{\mathrm{OSC}} + T_{\mathrm{IDLE\_PLL}}, \label{eq:uwb-start-latency} 
\end{equation}

and the consumed energy is

\begin{equation} 
    E_{\mathrm{start}}^{\mathrm{UWB}} = P_{\mathrm{OSC}}T_{\mathrm{OSC}} + P_{\mathrm{IDLE\_PLL}}T_{\mathrm{IDLE\_PLL}}. 
\label{eq:uwb-start-energy} 
\end{equation}

\subsubsection{Packet Airtime}
\begin{figure}
    \centering
    \includegraphics[width=1\linewidth]{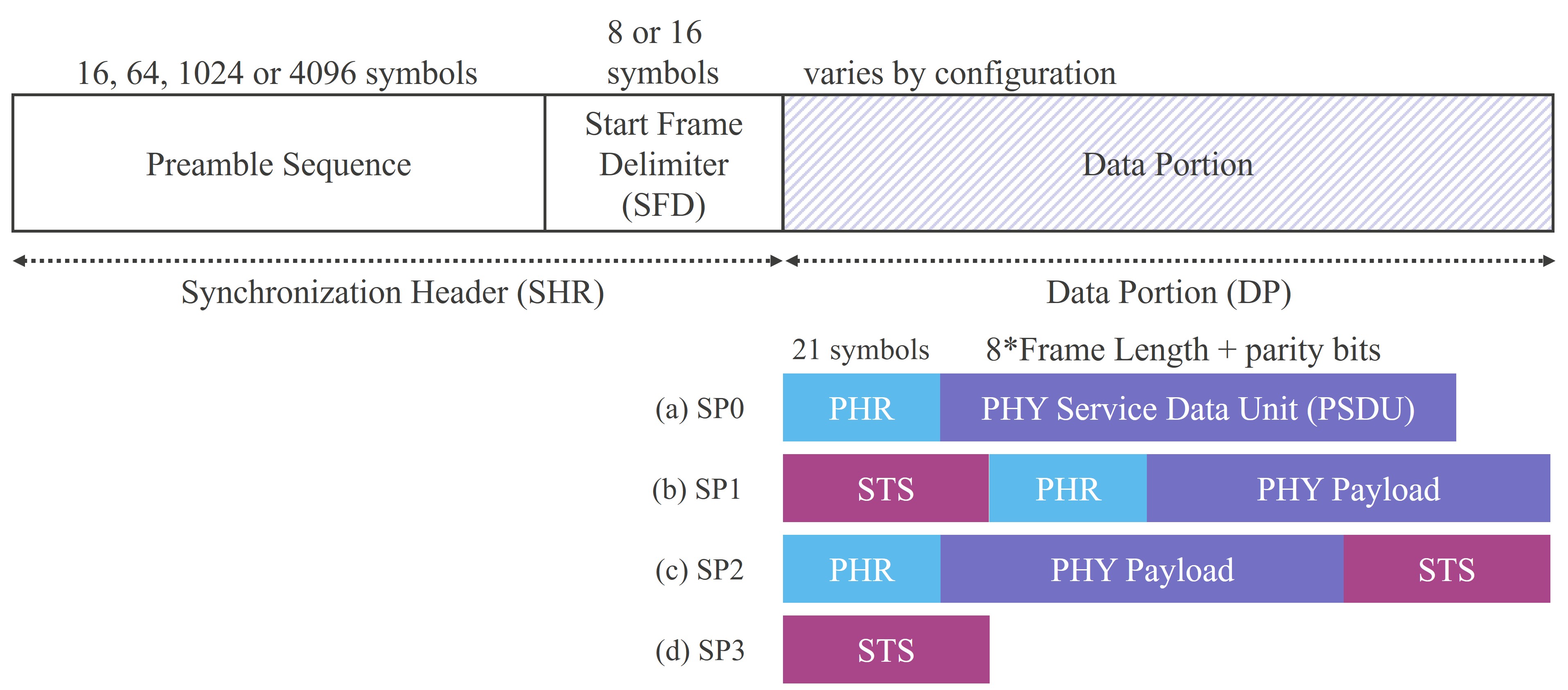}
    \caption{Packet Formats of IEEE802.15.4z \cite{coppens2022overview}.}
    \label{fig:ppdu-structure}
\end{figure}

In order to analyze the latency and power consumption of \gls{uwb} ranging, we consider the structure of the \gls{uwb} \gls{ppdu} (Fig. \ref{fig:ppdu-structure}), the symbol duration of Table \ref{tab:symbol-timings}, and the \gls{sts} packet configurations introduced by the IEEE802.15.4z amendment~\cite{coppens2022overview}.

Each \gls{ppdu} consists of a \gls{shr}, containing the preamble and \gls{sfd}, followed by a \gls{dp}. As shown in Fig. \ref{fig:ppdu-structure}, SP0 is the legacy configuration without an \gls{sts}, while SP1 and SP2 add the \gls{sts} before and after the payload, respectively. SP3 contains only the \gls{shr} and \gls{sts}, enabling timestamping without a \gls{phr} or payload. This version is used whenever a frame only needs to be timestamped, e.g. for the POLL and RESPONSE frames of a ranging exchange, while any additional information, such as the measured timestamps, is instead reported using one of the other frame formats.

Timing durations in IEEE 802.15.4/z are expressed as an integer number of symbols. Symbol durations depend on the data rate (0.85 or 6.81 Mbps for the DW3000) and the \acrfull{prf} (16 or 64 MHz). Table \ref{tab:symbol-timings} lists the resulting \gls{shr} symbol duration $\tau_{\mathrm{SHR}}$  and \gls{dp} symbol duration $\tau_{\mathrm{DATA}}$.

The duration of each field is then 
\begin{align}
    T_{\mathrm{SHR}} &= \tau_{\mathrm{SHR}}\left(N_{\mathrm{pre}} + N_{\mathrm{SFD}} + N_{\mathrm{STS}}\right),
    \label{eq:t-shr}\\
    T_{\mathrm{DATA}} &= \tau_{\mathrm{DATA}}\left(N_{\mathrm{PHR}} + 8L + 48\left\lceil \frac{8L}{330}\right\rceil\right),
    \label{eq:t-data}
\end{align}
where $N_{pre}$ is the number of preamble symbols, $N_{\mathrm{SFD}} \in \{8,16\}$, $N_{\mathrm{STS}}$ is the configured \gls{sts} length (a power of two between 32 and 2048 symbols, or $0$ if the packet configuration carries no \gls{sts}), $N_{\mathrm{PHR}} = 21$ symbols (or 0 in SP3), $L$ is the payload length in bytes ($L=0$ for SP3), and $48\left\lceil \frac{8L}{330}\right\rceil$ accounts for the Reed-Solomon parity bits appended to every 330 (or fewer) data bits.

Combining \eqref{eq:t-shr} and \eqref{eq:t-data}, the airtime of a packet
is simply
\begin{equation}
    T_{\mathrm{packet}} = T_{\mathrm{SHR}} + T_{\mathrm{DATA}}.
\label{eq:t-packet}
\end{equation}

\begin{table}
    \caption{DW3000 Symbol Duration \cite{dw3000usermanual}
    \label{tab:symbol-timings}
        }
\centering
\renewcommand{\arraystretch}{1.3}
\begin{tabular}{cccc}

\toprule
    \textbf{PRF (MHz)} & \textbf{Data Rate (Mbps)} & 
     \textbf{$\tau_{\mathrm{SHR}}$ (ns)} & \textbf{$\tau_{\mathrm{DATA}}$(ns)}\\
    \midrule
    
    16 & 0.85 & 993.59 & 1025.64 \\
    16 & 6.81 & 993.59 & 128.21 \\ 

    64 & 0.85 & 1017.63 & 1025.64 \\
    64 & 6.81 & 1017.63 & 128.21 \\
                           
\bottomrule
\end{tabular}
\end{table}

\subsubsection{Ranging Latency and Energy Consumption}
\begin{figure}[t]
    \centering
    \includegraphics[width=1\linewidth]{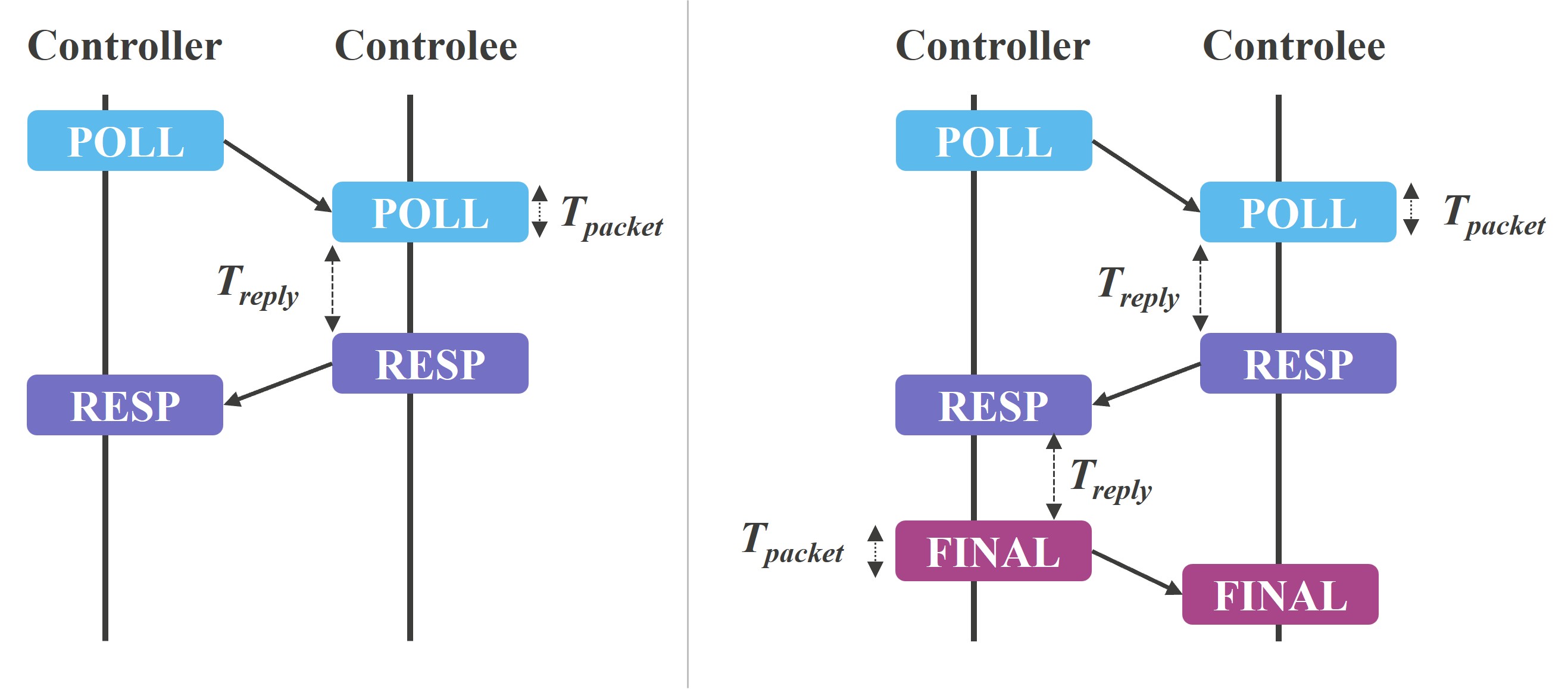}
    \caption{Timing diagram of \gls{twr}, \gls{sstwr} (left) and \gls{dstwr} (right).}
    \label{fig:TWR}
\end{figure}

The latency of a range round depends on the ranging mode, both modes are illustrated in Fig. \ref{fig:TWR}. \gls{sstwr} uses one request-reply exchange between two unsynchronized devices to measure the distance. \gls{dstwr} adds a second exchange in the opposite direction, combining both round-trip measurements to cancel out most of the clock-drift error.
For these two most common ranging modes, 
\begin{equation}
    n_f=
    \begin{cases}
    2, & \text{if ranging method }\text{\gls{ss}-\gls{twr} (Fig. \ref{fig:TWR} left)},\\
    3, & \text{else \gls{ds}-\gls{twr} (Fig. \ref{fig:TWR} right)}.
    \end{cases}
    \label{eq:nf}
\end{equation}

The airtime of every exchanged packet $T_{pkt,i}$ depends on the packet format, the payload and the reply time $T_{\mathrm{reply}}$ between reception and transmission.
\begin{equation} 
    T_{\mathrm{rng}} = \sum_{i=1}^{n_f} T_{\mathrm{pkt,i}} + (n_f - 1)T_{\mathrm{reply}}.
\label{eq:t-rng} 
\end{equation}

The energy consumed per ranging attempt depends on the airtime, ranging mode and role (transmit or receive) of every frame, which differs slightly depending on whether the device is the controller ($X=\mathrm{{ctrl}}$) or the controlee ($X=\mathrm{{cle}}$).

\begin{equation}
\begin{split}
    E_{\mathrm{rng}}^{X} ={}& P^{\mathrm{UWB}}_{\mathrm{tx}} \sum_{i \in \mathcal{F}^{X}_{\mathrm{tx}}} T_{\mathrm{pkt},i}
    + P^{\mathrm{UWB}}_{\mathrm{rx}} \sum_{i \in \mathcal{F}^{X}_{\mathrm{rx}}} T_{\mathrm{pkt},i} \\
    &+ (n_f - 1)\,T_{\mathrm{reply}}\,P_{\mathrm{IDLE\_PLL}},
\end{split}
\label{eq:energy-uwb-rng}
\end{equation}

where $\mathcal{F}^{\mathrm{ctrl}}_{\mathrm{tx}} = \{1, 3, 5, \dots\}$ and $\mathcal{F}^{\mathrm{ctrl}}_{\mathrm{rx}} = \{2, 4, 6, \dots\}$ denote the sets of frames transmitted and received by the controller, respectively, such that $|\mathcal{F}^{\mathrm{ctrl}}_{\mathrm{tx}}| = \lceil n_f / 2 \rceil$ and $|\mathcal{F}^{\mathrm{ctrl}}_{\mathrm{rx}}| = \lfloor n_f / 2 \rfloor$ (reversed for the controlee), and $P^{\mathrm{UWB}}_{\mathrm{tx}}$, $P^{\mathrm{UWB}}_{\mathrm{rx}}$, and $P_{\mathrm{IDLE\_PLL}}$ are the transmit, receive, and idle power levels. The parameters are listed in Table~\ref{tab:uwb_params}.

The total latency of the \gls{uwb} phase over $N_r$ ranging rounds is 
\begin{equation}
L_{\mathrm{UWB}} = N_r\,\left(T_{\mathrm{start}}^{\mathrm{UWB}} + T_{\mathrm{rng}}\right) + \left(N_r - 1\right)\,T_{\mathrm{deep\_sleep}}, 
\label{eq:latency-uwb-total} 
\end{equation} and the corresponding energy consumption is 

\begin{equation} 
\begin{split}
    E_{\mathrm{UWB}}^{X} &= N_r\,\left(E_{\mathrm{start}}^{\mathrm{UWB}} + E_{\mathrm{rng}}^{X}\right) \\&+ \left(N_r - 1\right)\,\left( P_{\mathrm{deep\_sleep}}T_{\mathrm{deep\_sleep}}\right).
\end{split}
\label{eq:energy-uwb-total} 
\end{equation}
\subsection{End-to-End Latency and Energy}
\label{subsec:end_to_end}
The session-level metrics compose the three phases: \gls{ble} discovery, \gls{ble} negotiation/commissioning, and \gls{uwb} ranging. Because the connection-oriented exchange of Section~\ref{subsec:neg} is only required when security material must be negotiated, the commissioning phase is present in connectable discovery ($m=\mathrm{conn}$) and absent in scannable discovery ($m=\mathrm{sc}$), where the configuration is delivered in the \texttt{SCAN\_RSP}. We capture this with the indicator $\mathbbm{1}_{\{m=\mathrm{conn}\}}$ equal to $1$ when its condition holds (i.e. $m=conn$) and $0$ otherwise.

\subsubsection{End-to-end Latency}
The session latency is measured from the instant the initiator starts scanning until the last ranging round completes. The discovery contribution is the scanner latency~\eqref{eq:expected_ble_latency_mode} and the post-discovery phases are common to both devices and independent of the phase offset $\phi$:
\begin{equation}
\mathbb{E}\!\left[L_{\mathrm{e2e}}^{(m)}\right] =
\mathbb{E}\!\left[L^{m}_{scan}\right]
+ \mathbbm{1}_{\{m=\mathrm{conn}\}}\, L_{\mathrm{neg}}
+ L_{\mathrm{UWB}} .
\label{eq:e2e_latency_scan}
\end{equation}

\subsubsection{End-to-end Energy}
Energy is accounted per device, as the two roles transmit and receive a different number of frames in every phase. We let the initiator/scanner act as \gls{ble} central / \gls{uwb} controller and the advertiser as \gls{ble} peripheral / \gls{uwb} controlee. Using the discovery energy~\eqref{eq:expected_energy_discovery}, the commissioning energy~\eqref{eq:neg-energy-central}, and the ranging energy~\eqref{eq:energy-uwb-total}:

\begin{equation}
    \begin{split}
    \mathbb{E}\!\left[E_{\mathrm{e2e}}^{\mathrm{cen},(m)}\right] =
    \mathbb{E}\!\left[E_{\mathrm{scan}}^{(m)}\right]
    &+ \mathbbm{1}_{\{m=\mathrm{conn}\}}\, E_{\mathrm{neg}}^{\mathrm{cen}}\\
    &+ E_{\mathrm{UWB}}^{\mathrm{ctrl}},
    \end{split}
    \label{eq:e2e:central_energy}
\end{equation}

\begin{equation}
    \begin{split}
    \mathbb{E}\!\left[E_{\mathrm{e2e}}^{\mathrm{per},(m)}\right] =
    \mathbb{E}\!\left[E_{\mathrm{adv}}^{(m)}\right]
    &+ \mathbbm{1}_{\{m=\mathrm{conn}\}}\, E_{\mathrm{neg}}^{\mathrm{per}} \\
    &+E_{\mathrm{UWB}}^{\mathrm{cle}},
    \end{split}
    \label{eq:e2e:peripheral_energy}
\end{equation}
\noindent

\section{Discussion and Model Scope}
\label{sec:discussions}
In this section, we use the analytical model to examine how \gls{ble} discovery and \gls{uwb} ranging influence the latency and energy consumption. 
Three discovery configurations are considered: low-latency, balanced, and low-power. The advertising intervals are inspired by Android settings, while the scan intervals and windows are selected to represent different duty cycles (100\%, 40\%, 10\%). These configurations and the remaining evaluation parameters are summarized in Table~\ref{tab:evaluation_parameters}. We consider connectable \gls{ble} discovery followed by parameter negotiation and \gls{dstwr} ranging. The scanner subsequently acts as the \gls{ble} central and \gls{uwb} controller, while the advertiser becomes the \gls{ble} peripheral and \gls{uwb} controlee.

\begin{table}[t]
\centering
\caption{Parameter values used in the discussion section.}
\centering \renewcommand{\arraystretch}{1.3}
\label{tab:evaluation_parameters}
\begin{tabular}{@{}lc@{}}
\hline
\textbf{Parameter} & \textbf{Selected value} \\
\hline
\multicolumn{2}{c}{\textit{\gls{ble} discovery configurations:
$T_{\mathrm{AI}}/T_{\mathrm{SI}}/T_{\mathrm{SW}}$}} \\
Low-latency & $100/100/100$ ms \\
Balanced & $250/500/200$ ms \\
Low-power & $1000/5000/500$ ms \\
\hline
\multicolumn{2}{c}{\textit{Common \gls{ble} settings}} \\
Discovery mode & Connectable \\
Advertising packet spacing $\tau_{\mathrm{wa}}$ & $1.5$ ms \\
ADV payload $S_{\mathrm{ADV}}$ & $37$ B \\
Connection interval $C$ & $7.5$ ms \\
Transmit window offset $O$ & $0$ ms \\
Transmit window size $W$ & $6.25$ ms \\
Session key size $S_{\mathrm{key}}$ & $16$ B \\
\hline
\multicolumn{2}{c}{\textit{Common UWB settings}} \\
Ranging method & \gls{dstwr} \\
\gls{prf} & $64$ MHz \\
Data rate & $6.81$ Mb/s \\
Preamble length $N_{\mathrm{pre}}$ & $64$ symbols \\
\gls{sfd} length $N_{\mathrm{SFD}}$ & $8$ symbols \\
\gls{sts} length $N_{\mathrm{STS}}$ & $64$ symbols \\
POLL/RESPONSE packet format & SP3 \\
FINAL packet format & SP1 \\
FINAL payload length $L$ & $12$ B \\
Inter-frame reply time $T_{\mathrm{reply}}$ & $400\,\mu$s \\
Inter-round sleep time $T_{\mathrm{deep\_sleep}}$ & $100$ ms \\
\hline
\end{tabular}
\end{table}

\subsection{Energy Costs of the Scanner and the Advertiser Roles}
\label{sub:roles}

\begin{figure*}
     \centering
         \subfloat[]{\includegraphics[width=0.48\linewidth]{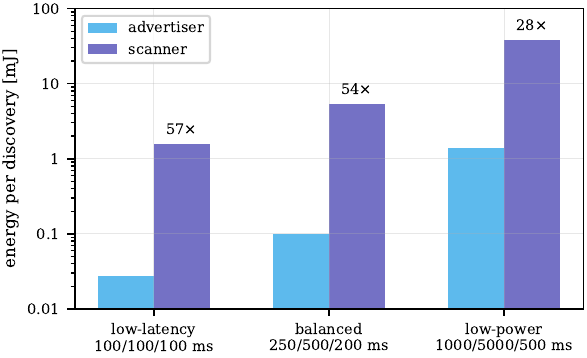}
         \label{fig:discovery_energy}}
     \hfill
         \subfloat[]{\includegraphics[width=0.48\linewidth]{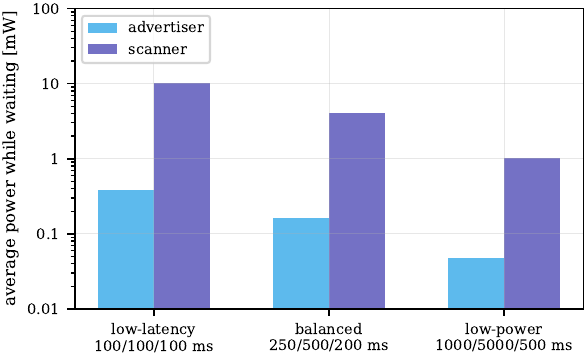}
         \label{fig:waiting_power}}
    \caption{Energy consumption of advertiser and scanner discovery for the low-latency, balanced, and low-power configurations ($T_{\mathrm{AI}}$/$T_{\mathrm{SI}}$/$T_{\mathrm{SW}}$): (a) average energy per discovery event and (b) average power consumed while waiting for a discovery opportunity. Lower duty cycles reduce waiting power but increase energy cost of a successful discovery due to longer discovery latencies.}
    \label{fig:discovery_energy_characteristics}
\end{figure*}

Figure \ref{fig:discovery_energy_characteristics} shows the expected energy required to complete a discovery event for both the advertiser and scanner roles. Across the three parameter configurations, the scanner consumes more energy than the advertiser, requiring approximately 28$\times$, 54$\times$, and 57$\times$ more energy. This difference arises because the scanner spends significantly longer with the receiver active, whereas advertising consists of short transmission bursts.
While Fig.~\ref{fig:discovery_energy} quantifies the energy required to complete a discovery event, Fig~\ref{fig:waiting_power} shows the average power consumed while waiting for a discovery opportunity. The latter is obtained as
\begin{equation} 
P_{\mathrm{adv,wait}} = \frac{E_{\mathrm{adv,no}}} {\bar{T}_{\mathrm{adv}}} ,
\qquad
P_{\mathrm{scan,wait}} = \frac{E_{\mathrm{scan,no}}} {T_{\mathrm{SI}}} .
\label{eq:average_waiting}
\end{equation}

As shown in Fig. \ref{fig:waiting_power}, the average power consumed while waiting decreases as the duty cycle is reduced. For the scanner, the average waiting power drops from approximately 10.2 mW in the low-latency configuration to 1.03 mW in the low-power configuration, reflecting the reduced time spent actively listening. Although the low-power configuration consumes less power over time, it also increases the expected discovery latency because the scanner listens less frequently. The longer waiting time before a discovery occurs outweighs the reduction in instantaneous power consumption, leading to a higher total energy cost per discovery. For example, the scanner discovery energy increases from approximately 1.6 mJ in the low-latency configuration to 38.2 mJ in the low-power configuration. These results highlight the fundamental duty-cycling trade-off. Reducing the duty cycle lowers the average power consumption while waiting for a discovery event, but increases the expected discovery latency.


\subsection{Discovery Latency Analysis}
\label{sub:discovery_latency_distribution}

\begin{figure}
  \centering
  \includegraphics[width=\columnwidth]{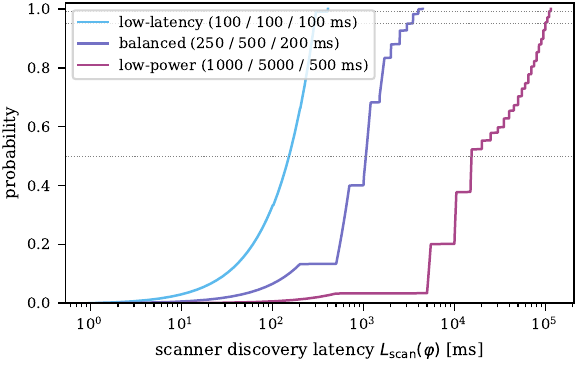}
  \caption{Cumulative distribution of scanner discovery latency over the phase offset $\phi$ for three \gls{ble} configurations $(T_{\mathrm{AI}}/T_{\mathrm{SI}}/T_{\mathrm{SW}})$.}
  \label{fig:latency-distribution}
\end{figure}

The expected discovery latency in \eqref{eq:expected_ble_latency_mode} describes only the average behavior. For interactive applications such as digital keys, interactive museum displays, collision warnings, and access control, however, higher-percentile latencies are often more relevant than the mean. Fig.~\ref{fig:latency-distribution} shows the cumulative distribution of the scanner discovery latency for the same three \gls{ble} configurations. All distributions exhibit a right tail, with p99 latencies ranging from 2$\times$ to 3.4$\times$ the mean latency. The low-latency configuration achieves a mean latency of 153~ms and a p99 latency of 302~ms. For the balanced configuration, these values increase to 1.19 s and 4.01 s, respectively, while the low-power configuration yields a mean of 35.8 s and a p99 of 110 s. These results demonstrate that the mean latency alone does not fully capture discovery performance. As the duty cycle decreases, the tail of the latency distribution grows significantly, making percentile metrics such as p95 and p99 essential for assessing the responsiveness experienced by users. These large latencies may be acceptable for infrequent asset-tracking applications, but are unsuitable for interactive use cases such as digital keys, access control, or point-and-control interactions.

\subsection{Discovery Energy versus Ranging Energy}
\label{sub:discovery_ranging_crossover}

The total energy of a ranging session can be decomposed into a fixed setup cost, comprising \gls{ble} discovery and commissioning, and a ranging-dependent cost due to \gls{uwb} startup and \gls{dstwr} exchanges. The crossover point $N_r^\star$ is defined as the number of ranging rounds at which the cumulative \gls{uwb} ranging energy equals the one-time energy associated with session establishment.

\begin{equation} 
    N_r^{\star} = \left\lceil\frac{\mathbb{E}\!\left[E_{x}^{(m)}\right] + \mathbbm{1}_{\{m=\mathrm{conn}\}}\,E_{\mathrm{neg}}^{X} + P_{\mathrm{deep\_sleep}}T_{\mathrm{deep\_sleep}}}{E_{\mathrm{start}}^{\mathrm{UWB}} + E_{\mathrm{rng}}^{X} + P_{\mathrm{deep\_sleep}}T_{\mathrm{deep\_sleep}}}\right\rceil
\end{equation}

\begin{figure}
  \centering
  \includegraphics[width=\columnwidth]{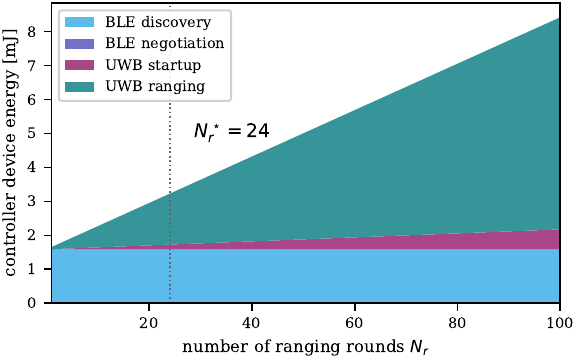}
  \caption{Energy breakdown of the controller device as a function of the number of \gls{dstwr} ranging rounds for $T_{\mathrm{AI}} = T_{\mathrm{SI}} = T_{\mathrm{SW}} = 100$ ms. \gls{ble} discovery and commissioning contribute a fixed one-time energy overhead, whereas \gls{uwb} startup and ranging energy increase with the number of ranging rounds. The crossover point $N_r^\star$ denotes the number of ranging rounds at which the cumulative \gls{uwb} energy equals the \gls{ble} session-establishment energy.}
  \label{fig:budget}
\end{figure}

Figure \ref{fig:budget} illustrates this crossover point for $T_{\mathrm{AI}} = 100$ ms, $T_{\mathrm{SI}} = 100$ ms, $T_{\mathrm{SW}} = 100$ ms. The figure shows the energy consumption of the scanning device, which subsequently assumes the \gls{ble} central role and the \gls{uwb} controller role. 

For this configuration, the crossover occurs at $N_r^\star = 24$. Therefore, after 24 complete \gls{dstwr} ranging rounds, the cumulative energy consumed by \gls{uwb} operation exceeds the one-time energy required to establish the session through \gls{ble} discovery and commissioning. For shorter sessions, the \gls{ble} initialization phase contributes a significant fraction of the total energy consumption. As the number of ranging rounds increases, however, the impact of the \gls{ble} overhead becomes less important and the total energy is increasingly determined by \gls{uwb} operation. 

Although Fig. \ref{fig:budget} includes the negotiation energy associated with connectable operation, its contribution is negligible compared to the discovery energy. For the considered configuration, the negotiation phase consumes approximately 11.2 $\mathrm{\mu}$J, whereas \gls{ble} discovery requires approximately 1.6 mJ. Consequently, the negotiation contribution is effectively invisible at the scale of the figure. A purely scannable implementation without \gls{ble} connection establishment would therefore exhibit nearly identical crossover behavior in the ranging case. 

The value of $N_r^\star$ depends directly on the energy consumed during the \gls{ble} discovery phase. Configurations with lower discovery energy require fewer ranging rounds before \gls{uwb} becomes the dominant contributor to the total energy budget. For the low-latency configuration shown in Fig. \ref{fig:budget}, the crossover occurs after only 24 ranging rounds. In comparison, the balanced configuration and the low-power configuration yield crossover points of approximately 80 and 562 rounds, respectively. These results demonstrate that the relative importance of \gls{ble} discovery and UWB ranging is highly dependent on the selected discovery parameters.

The crossover analysis highlights the complementary roles of \gls{ble} and \gls{uwb} in FiRa-like systems. \gls{ble} introduces a one-time discovery overhead that enables devices to exchange the information required to initiate a ranging session, while \gls{uwb} provides the actual ranging functionality. For short-lived sessions involving only a limited number of ranging exchanges, \gls{ble} discovery can account for a substantial portion of the total energy consumption. For longer sessions, the initialization overhead becomes amortized over a large number of ranging rounds and the total energy budget is dominated by \gls{uwb} operation.

\section{From Ranging to Localization}
\label{sec:ranging-to-localization}
The model so far assumed a device-to-device session ending in a \gls{twr} exchange. In an infrastructure deployment the tag is instead localized by a group of anchors, using one of three \gls{uwb} localization approaches: \gls{twr} with each anchor in turn; \gls{ultdoa}, in which the \gls{uwb} tag transmits and the synchronized anchors report their received timestamps to a server; or \gls{dltdoa}, in which the\gls{uwb} anchors transmit and the tag solves for its own position. These architectures can be expanded to estimate the \gls{aoa} in multi-antenna systems.
Since the \gls{ble} parameters a tag needs to join a zone are identical for every tag in that zone, our model delivers them in the \texttt{SCAN\_RSP}. This approach minimizes complexity and is sufficient whenever neither authentication nor parameter negotiation is required. Some deployments, however, require device authentication, credential exchange, or dynamic negotiation of parameters. In such cases, connectable advertisements are used to establish a \gls{ble} connection. The resulting connection enables secure transfer and configuration at the cost of additional discovery latency and higher \gls{ble} energy consumption.

In this section, we investigate how discovery behaves against a group of advertisers rather than a single one, and how one \gls{uwb} round changes when ranging is replaced by localization. Because the anchors are typically mains-powered, the energy model considers only the tag.
 
\subsection{Discovery of an Anchor Group}
Anchors of the same group advertise the same infrastructure identity, so the tag does not discover each anchor individually: discovery ends as soon as any anchor of the group is discovered. Let $N_\mathrm{A}$ be the number of anchors in the group and $N_\mathrm{A}^{\mathrm{vis}} \le N_\mathrm{A}$ the number of anchors within range of the tag. Eqs.~\eqref{eq:success_cond}--\eqref{eq:ble_latency_scan} hold per anchor $j$ with its own offset $\phi_j \sim \mathcal{U}[0,3T_{\mathrm{SI}})$, and the anchors are mutually unsynchronized, so the different $\phi_j$ are independent and identically distributed. The group discovery latency is then
\begin{equation}
  L_{\mathrm{scan}}^{\mathrm{loc}}(\boldsymbol{\phi})
  = \min_{j \le N_\mathrm{A}^{\mathrm{vis}}} L_{\mathrm{scan}}(\phi_j),
  \qquad \boldsymbol{\phi}=(\phi_1,\dots,\phi_{N_\mathrm{A}^{\mathrm{vis}}}).
\end{equation}
Using the identity $\mathbb{E}[X]=\int_0^\infty \Pr(X>t)\,\mathrm{d}t$ for a non-negative random variable $X$, with $t$ ranging over all candidate discovery times, the single integral of
\eqref{eq:expected_ble_latency} is replaced by a product of complementary distribution functions,
\begin{equation}
\begin{split}
    \mathbb{E}\!\left[L_{\mathrm{scan}}^{\mathrm{loc}}\right]
  &= \int_0^{\infty} \bar{F}(t)^{\,N_\mathrm{A}^{\mathrm{vis}}}\,\mathrm{d}t,
  \quad\\
  \bar{F}(t)&=\frac{\bigl|\{\phi : L_{\mathrm{scan}}(\phi) > t\}\bigr|}{3T_{\mathrm{SI}}},
\end{split}
\end{equation}
which is strictly decreasing in $N_\mathrm{A}^{\mathrm{vis}}$ where a dense deployment is discovered faster than a single peer at the same advertising interval (see Section \ref{sub:discussion-localization}).
\subsection{UWB Phase}
A single distance is no longer enough to calculate the position of a tag; at least three anchors are needed in 2D and four in 3D~\cite{tiemann2019scalability}. We assume a coordinated infrastructure in which discovery of any anchor provides the group configuration, or a reference to a predefined profile, including the information needed to identify the participating anchors and their scheduled exchanges. We further assume that the anchors visible over \gls{ble} are also within \gls{uwb} range, and that the anchors are synchronized for both \gls{tdoa} modes. The establishment of the schedule and the synchronization are outside the scope of the model. 

The three localization schemes differ in the number and direction of their frame exchanges. In \gls{twr}, the tag ranges with $N_\mathrm{A}^{\mathrm{use}} \le N_\mathrm{A}^{\mathrm{vis}}$ anchors and repeats \eqref{eq:t-rng} once per anchor. Since every additional anchor costs a full exchange, the tag ranges with only as many anchors as required, i.e. at least three in 2D and four in 3D. In \gls{ultdoa}, the tag transmits a single uplink frame, which every visible anchor receives and timestamps, so all $N_\mathrm{A}^{\mathrm{vis}}$ anchors contribute at no extra cost to the tag. \gls{dltdoa} reverses this: the anchors transmit a block of $N_\mathrm{A}^{\mathrm{vis}}$ frames in assigned slots, and the tag receives all of them.

We capture all schemes $Y \in \{\mathrm{SS},\mathrm{DS},\mathrm{UL},\mathrm{DL}\}$ by letting the frame count $n_f^{Y}$, the time between frames $T_{\mathrm{reply}}^{Y}$, and the number of exchanges per position fix $N_{\mathrm{ex}}^{Y}$ depend on the scheme:
\begin{equation}
  T_{\mathrm{rng}}^{Y} = \sum_{i=1}^{n_f^{Y}} T_{\mathrm{pkt},i}
  + \bigl(n_f^{Y}-1\bigr)\,T_{\mathrm{reply}}^{Y},
  \label{eq:duration_one_round}
\end{equation}
\begin{equation}
  T_{\mathrm{fix}}^{Y} = N_{\mathrm{ex}}^{Y}\,T_{\mathrm{rng}}^{Y} + \bigl(N_{\mathrm{ex}}^{Y} - 1\bigr)\,T_{\mathrm{IDLE\_PLL}}.
  \label{eq:duration_more_rounds}
\end{equation}
Equation~\eqref{eq:duration_one_round} gives the duration of one exchange, and \eqref{eq:duration_more_rounds} repeats it $N_{\mathrm{ex}}^{Y}$ times, separated by an idle time $T_{\mathrm{IDLE\_PLL}}$, to obtain one position estimate. Table~\ref{tab:uwb_localization} summarizes the parameters of each scheme.

\begin{table}[t]
\centering
\renewcommand{\arraystretch}{1.3}
\caption{Parameters for Localization Schemes (Tag Side)}
\label{tab:uwb_localization}

\begin{tabular}{lccccc}
\toprule
\textbf{Scheme} & $n_f$ & $n_{\mathrm{tx}}$ & $n_{\mathrm{rx}}$ & $T_{\mathrm{reply}}$ & $N_{\mathrm{ex}}$ \\
\midrule
\gls{sstwr}  & 2       & 1 & 1       & $T_{\mathrm{reply}}$ & $N_A^{\mathrm{use}}$ \\
\gls{dstwr}  & 3       & 2 & 1       & $T_{\mathrm{reply}}$ & $N_A^{\mathrm{use}}$ \\
\gls{ultdoa} & 1       & 1 & 0       & --                   & 1 \\
\gls{dltdoa} & $N_A^{\mathrm{vis}}$ & 0 & $N_A^{\mathrm{vis}}$ & $T_{\mathrm{slot}}$  & 1 \\
\bottomrule
\end{tabular}
\end{table}

Generalizing \eqref{eq:latency-uwb-total}, the total \gls{uwb} phase latency for localization is
\begin{equation}
\begin{split}
    L_{\mathrm{UWB}}^{Y} &= N_r\bigl(T_{\mathrm{start}}^{\mathrm{UWB}}
  + T_{\mathrm{fix}}^{Y}\bigr)
  + (N_r-1)\,T_{\mathrm{deep\_sleep}}\\
  &+ \mathbbm{1}_{\{Y=\mathrm{UL}\}} T_{\mathrm{infra}},
  \label{eq:uwb_latency_localization}
\end{split}
\end{equation}
where $T_{\mathrm{infra}}$ covers the anchor-to-server reporting, the multilateration, and, if the tag needs to know its own position, the downlink delivery of the result. It is only added for $Y=\mathrm{UL}$, since only in \gls{ultdoa} is the position computed by the infrastructure. In \gls{twr} and \gls{dltdoa}, the tag computes its own position.

The per-exchange energy of \eqref{eq:energy-uwb-rng} generalizes to
\begin{equation}
\begin{split}
  E_{\mathrm{rng}}^{Y} ={}& P^{\mathrm{UWB}}_{\mathrm{tx}} \sum_{i \in \mathcal{F}^{Y}_{\mathrm{tx}}} T_{\mathrm{pkt},i}
  + P^{\mathrm{UWB}}_{\mathrm{rx}} \sum_{i \in \mathcal{F}^{Y}_{\mathrm{rx}}} T_{\mathrm{pkt},i}\\
  &+ \bigl(n_f^{Y}-1\bigr)\,T_{\mathrm{reply}}^{Y}\,P_{\mathrm{IDLE\_PLL}},
\end{split}
\label{eq:uwb_energy_localization}
\end{equation}
where $\mathcal{F}^{Y}_{\mathrm{tx}}$ and $\mathcal{F}^{Y}_{\mathrm{rx}}$ denote the sets of frames transmitted and received by the tag, with $|\mathcal{F}^{Y}_{\mathrm{tx}}| = n^{Y}_{\mathrm{tx}}$ and $|\mathcal{F}^{Y}_{\mathrm{rx}}| = n^{Y}_{\mathrm{rx}}$ as listed in Table~\ref{tab:uwb_localization}. The energy per position fix is
\begin{equation}
  E^{Y}_{\mathrm{fix}}
  = N^{Y}_{\mathrm{ex}}\,E^{Y}_{\mathrm{rng}}
  + \bigl(N^{Y}_{\mathrm{ex}}-1\bigr)\,P_{\mathrm{IDLE\_PLL}}\,T_{\mathrm{IDLE\_PLL}},
  \label{eq:E_fix}
\end{equation}
and the total \gls{uwb} phase energy over $N_r$ rounds is
\begin{equation}
\begin{split}
  E^{Y}_{\mathrm{UWB}}
  ={}& N_{r}\bigl(E_{\mathrm{start}}^{\mathrm{UWB}} + E^{Y}_{\mathrm{fix}}\bigr)\\
  &+ (N_{r}-1)\,P_{\mathrm{deep\_sleep}}\,T_{\mathrm{deep\_sleep}}.
\end{split}
\label{eq:E_UWB_loc}
\end{equation}

The two \gls{tdoa} modes differ markedly in tag energy. In \gls{ultdoa}, the tag only transmits a single frame, so its energy is independent of the number of anchors:
\begin{equation}
  E_{\mathrm{rng}}^{\mathrm{UL}} = T_{\mathrm{pkt}}\,P^{\mathrm{UWB}}_{\mathrm{tx}}.
  \label{eq:E_UL}
\end{equation}
In \gls{dltdoa}, the tag does the reverse and receives all $N_\mathrm{A}^{\mathrm{vis}}$ anchor frames, which, for frames of identical airtime $T_{\mathrm{pkt}}$, gives
\begin{equation}
  E_{\mathrm{rng}}^{\mathrm{DL}} = N_\mathrm{A}^{\mathrm{vis}}\,T_{\mathrm{pkt}}\,P^{\mathrm{UWB}}_{\mathrm{rx}} + \bigl(N_\mathrm{A}^{\mathrm{vis}}-1\bigr)\,T_{\mathrm{slot}}\,P_{\mathrm{IDLE\_PLL}}.
  \label{eq:E_DL}
\end{equation}

\subsection{Discussion}
\label{sub:discussion-localization}

\begin{figure*}
     \centering
         \subfloat[]{\includegraphics[width=0.48\linewidth]{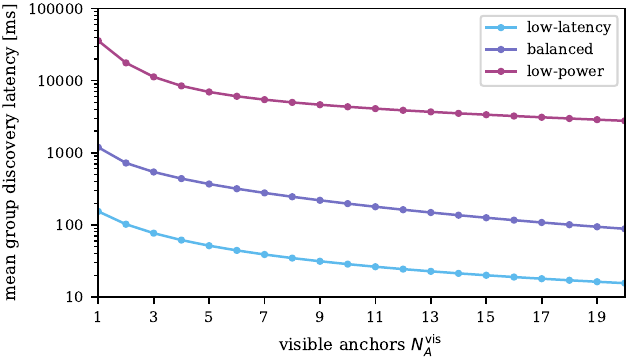}
         \label{fig:group_discovery_latency}}
     \hfill
         \subfloat[]{\includegraphics[width=0.48\linewidth]{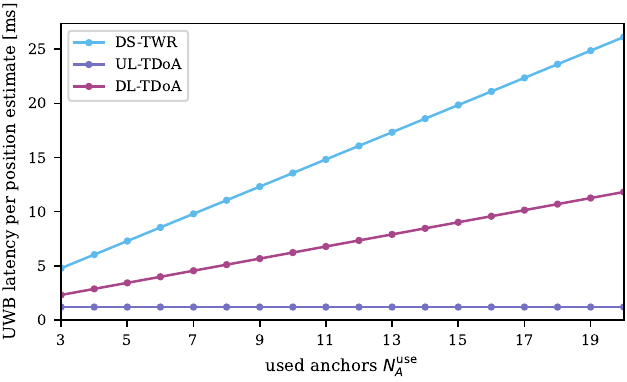}
         \label{fig:uwb_localization_latency}}
    \caption{(a) Expected latency (logarithmic scale) to discover any one anchor versus the number of visible anchors $N_\mathrm{A}^{\mathrm{vis}}$ for three \gls{ble} configurations. (b) \gls{uwb} latency per position estimate on the tag side versus the number of used anchors $N_A^{\mathrm{use}}$ for \gls{dstwr}, \gls{ultdoa}, and \gls{dltdoa}.}
    \label{fig:localization_latency}
\end{figure*}

Fig~\ref{fig:localization_latency} provides an analysis of the latency of the \gls{ble} anchor discovery phase and the \gls{uwb} localization phase. Fig.~\ref{fig:group_discovery_latency} shows that the expected group discovery latency decreases with $N_\mathrm{A}^{\mathrm{vis}}$ for all three \gls{ble} configurations, since discovery completes at the earliest of $N_\mathrm{A}^{\mathrm{vis}}$ independent advertisers. The largest gains occur for the first few anchors: three visible anchors already halve the latency compared to a single one, from 154 ms to 77 ms for the low-latency configuration and from 35.8 s to 11.3 s for the low-power configuration. With twenty visible anchors, the latency is reduced by a factor of 10 to 13 across all configurations, down to 16 ms, 88 ms, and 2.8 s, respectively. A dense deployment therefore allows longer advertising intervals, and thus lower anchor duty cycles, for the same discovery latency. Since anchors are often mains powered, potentially even lower advertisement intervals can be used to further reduce the discovery latency.

Fig.~\ref{fig:uwb_localization_latency} shows the \gls{uwb} latency per position estimate on the tag side, i.e. excluding $T_{\mathrm{infra}}$. For \gls{dstwr}, the latency grows linearly with $N_\mathrm{A}^{\mathrm{use}}$ by 1.26 ms per anchor, from 4.8 ms for three anchors to 26.1 ms for twenty, since every additional anchor adds a full three-frame exchange and an idle gap. Thus, \gls{twr} provides high ranging accuracy, requires no network synchronization, and relies on mature, well-understood protocols, making it well suited for secure access control and point-to-point positioning. However, if secure localization is required, \gls{ble} discovery and negotiation must be performed with each participating anchor, and scalability is limited because every tag must actively exchange messages with every anchor it uses.

\gls{dltdoa} also grows linearly, but by only 0.56 ms per anchor, since the tag must remain awake until it has received the frames of all participating anchors, each requiring a single frame and slot instead of a full exchange. Thus, \gls{dltdoa} offers strong privacy, as the tag computes its position locally without reporting measurements to the infrastructure. It is therefore particularly attractive for smartphone navigation and user-centric localization services. However, the tag must receive frames from all visible anchors and maintain its clock reference between receptions, which prevents the use of (deep) sleep modes during the block and makes its energy grow with $N_\mathrm{A}^{\mathrm{vis}}$.

The latency of \gls{ultdoa} is constant at 1.2 ms, since the tag transmits a single frame regardless of the number of anchors. In \gls{ultdoa}, therefore, a larger number of anchors improves the position estimate at no cost in tag latency. Thus, \gls{ultdoa} minimizes the tag's communication overhead and energy consumption, and since the tag's airtime is independent of the number of anchors, it scales well to many tags and anchors. This makes it attractive for low-power, large-scale industrial \gls{rtls} applications such as asset and worker tracking. Its main drawback is that the position estimate resides within the infrastructure rather than on the device itself.

The choice of architecture is primarily driven by scalability, privacy requirements, and whether the position estimate is required on the device or within the infrastructure. \gls{ble} serves as a common admission mechanism across all approaches, enabling infrastructure discovery, parameter distribution, and optional authentication while leveraging the \gls{ble} and \gls{uwb} radios already integrated in many smartphones and \acrshort{iot} devices.

\section{Limitations and Future Work}
\label{sec:limitations-future-work}
Several extensions could further increase the accuracy of the proposed model. First, the current analysis considers a single advertiser-scanner pair and, for localization, a single tag interacting with independent anchors on an otherwise idle medium. Future work could investigate the impact of network density, contention, collisions, and anchor scheduling on both discovery performance and energy consumption.

Second, the \gls{ble} discovery model assumes that the radio can enter a sleep state between consecutive scan and advertising windows. In practice, the achievable power state depends on the ratio between the configured interval and window durations, and short gaps may require the radio to remain in an idle or standby state. Incorporating adaptive power-state selection would improve the accuracy of the energy model.

Finally, following \cite{shih2022latency}, the discovery analysis employs the mean-delay approximation of the \gls{ble} advertising delay. While this approximation is accurate for the non-resonant parameter configurations considered in this work, future models could explicitly incorporate the random advertising delay to study resonant operating points and their impact on discovery latency.

\section{Conclusion}
\label{sec:conclusion}
This paper presented an analytical model of the end-to-end latency and energy consumption of \gls{ble}-assisted \gls{uwb} ranging and localization systems. By combining \gls{ble} discovery, \gls{ble} negotiation, \gls{uwb} startup and ranging in a single framework, the model quantifies the cost of an entire session and attributes it to individual phases and device roles. The analysis leads to three main observations. First, \gls{ble} discovery is highly asymmetric: depending on the discovery configuration, the scanner consumes between 28$\times$ and 57$\times$ more energy than the advertiser because it must keep its receiver active while the advertiser transmits only short packets. Second, reducing the \gls{ble} duty cycle lowers the average power while waiting for discovery, but increases both the energy required per successful discovery and the tail of the latency distribution. Across the evaluated configurations, the 99th-percentile discovery latency ranges from 2$\times$ to 3.4$\times$ the mean, reaching up to 110~s. Third, the energy required to establish the \gls{ble} connection and negotiate session parameters is negligible compared to the discovery phase, and the associated setup overhead is amortized only after 24-562 ranging rounds. For localization, the model highlights the different scaling behavior of the considered architectures. \gls{ultdoa} keeps the tag-side energy consumption independent of the number of anchors, whereas \gls{dltdoa} and \gls{twr} require additional tag activity as anchor count increases. The results also show that group discovery becomes more efficient as more anchors are available, reducing the expected time until a suitable anchor is detected. Overall, the results indicate that \gls{ble} discovery should not be treated as a negligible bootstrap step. In many practical deployments, particularly those involving short interactions, discovery dominates the latency and energy budget of the complete \gls{uwb} session. Consequently, the choice of discovery mode, parameter configuration, and device role should be considered an integral part of the design of \gls{ble}-assisted \gls{uwb} ranging and localization systems.

\section*{Acknowledgments}
This research was funded by the imec ICON project NG-UWB (Agentschap Innoveren en Ondernemen project nr. HBC.2024.0882) and the DistriMuSe project (HORIZON-KDT-JU-2023-2-RIA HORIZON JU Research and Innovation Actions) under Grant 101139769.

The authors used Microsoft Copilot and Claude (Anthropic) for language editing and for assistance with the formulation of the text and equations, and Claude to assist with writing the code. All AI-assisted text, equations, and code were reviewed and validated by the authors, who take full responsibility for the content of this article.

\bibliographystyle{IEEEtran}
\bibliography{references.bib}

\vfill

\end{document}